%% file: cas-dc-template.tex
\documentclass[a4paper,fleqn]{cas-dc}

\usepackage[authoryear]{natbib}
\usepackage{listings}					% For listing any code in LaTeX
\usepackage{enumerate}					% For making list (IMPORTANT)
\usepackage{tabularx}					% For making fancy tables (IMPORTANT)
\usepackage[normalem]{ulem}				% For highlighting and emphasis
\usepackage[flushleft]{threeparttable}		% For miscellaneous alignments
\usepackage{changepage}	% For \adjustwidth command for writing pseudocodes
\usepackage{url}
\usepackage{caption}
\usepackage{amsmath}
\usepackage{algpseudocode}
\usepackage{float}
\usepackage{cleveref}
\usepackage{amsmath,amssymb,amsfonts}
\usepackage{graphicx}
\usepackage{textcomp}
\usepackage{xcolor}
\usepackage{multirow}
\usepackage{float}
\usepackage{eurosym}
\usepackage{booktabs}
\usepackage{subfig}
\usepackage{svg}
\definecolor{gray}{rgb}{.949,.949,.949}
\newcolumntype{Y}{>{\centering\arraybackslash}X}
\hypersetup{hidelinks,colorlinks=false,pdfstartview=Fit,breaklinks=true}
\def\tsc#1{\csdef{#1}{\textsc{\lowercase{#1}}\xspace}}
\tsc{WGM}
\tsc{QE}
\tsc{EP}
\tsc{PMS}
\tsc{BEC}
\tsc{DE}
\begin{document}
\let\WriteBookmarks\relax
\def\floatpagepagefraction{1}
\def\textpagefraction{.001}

\shorttitle{GAIA - A Large Language Model for Advanced Power Dispatch}

\shortauthors{Yuheng Cheng et~al.}

\title [mode = title]{GAIA - A Large Language Model for Advanced Power Dispatch}                      
% \tnotemark[1,2]


% TODO
% \tnotetext[1]{This work was partially supported by the National Natural Science Foundation of China under Grant 7217010719.}

% \tnotetext[2]{This work was supported in part by the Shenzhen Institute of Artificial Intelligence and Robotics for Society under Grant AC01202101102.}

\author[1,2]{Yuheng Cheng}[style=chinese]
\cormark[1]
% \cormark[1]


\ead{yuhengcheng@link.cuhk.edu.cn}

\author[3]{Huan Zhao}[style=chinese]
\cormark[1]
\ead{huan-paul.zhao@polyu.edu.hk}

% \ead[url]{jiaqiruan@link.cuhk.edu.cn}

% \credit{Conceptualization of this study, Methodology, Software}


\author[4]{Xiyuan Zhou}[style=chinese]
% \cormark[2]
\ead{xiyuan002@e.ntu.edu.sg}

\author[1,2]{Junhua Zhao}[style=chinese]
\cormark[2]
\ead{zhaojunhua@cuhk.edu.cn}

\author[5]{Yuji Cao}[style=chinese]
% \cormark[2]
\ead{yujicao@link.cuhk.edu.cn}

\author[6]{Chao Yang}[style=chinese]
% \cormark[2]
\ead{yangchao@ncepu.edu.cn)}

% \credit{Data curation, Writing - Original draft preparation}
    
\affiliation[1]{organization={Shenzhen Institute of Artificial Intelligence and Robotics for Society (AIRS)},
    city={Shenzhen},
    % citysep={}, % Uncomment if no comma needed between city and postcode
    postcode={518129}, 
    country={China}}
    
\affiliation[2]{organization={School of Science and Engineering, The Chinese University of Hong Kong},
    % addressline={Radarweg 29}, 
    city={Shenzhen},
    % citysep={}, % Uncomment if no comma needed between city and postcode
    postcode={518172}, 
    country={China}}

\affiliation[3]{organization={Department of Building Environment and Energy Engineering, The Hong Kong Polytechnic University},
    % addressline={Radarweg 29}, 
    city={Hong Kong},
    % citysep={}, % Uncomment if no comma needed between city and postcode
    %postcode={}, 
    country={China}}

\affiliation[4]{organization={School of Electrical and Electronic Engineering, Nanyang Technological University},
    % addressline={Radarweg 29}, 
    city={Singapore},
    % citysep={}, % Uncomment if no comma needed between city and postcode
    postcode={639798}, 
    country={Singapore}}

\affiliation[5]{organization={Department of Mechanical and Automation Engineering, The Chinese University of Hong Kong},
    % addressline={Radarweg 29},
    city={Hong Kong},
    % citysep={}, % Uncomment if no comma needed between city and postcode
    postcode={999077}, 
    country={China}}

\affiliation[6]{organization={School of Electrical and Electronic Engineering, North China Electric Power University},
    % addressline={Radarweg 29}, 
    city={Baoding},
    % citysep={}, % Uncomment if no comma needed between city and postcode
    postcode={071003}, 
    country={China}}
    
\cortext[cor1]{These authors contributed to the work equally and should be regarded as co-first authors.}
\cortext[cor2]{Corresponding author at: School of Science and Engineering, The Chinese University of Hong Kong, Shenzhen, 518172, China.}

% \cortext[cor2]{Principal corresponding author}

% \fntext[fn1]{This is the first author footnote. but is common to third
% \fntext[fn2]{Another author footnote, this is a very long footnote and
%   it should be a really long footnote. But this footnote is not yet
%   sufficiently long enough to make two lines of footnote text.}

% \nonumnote{This note has no numbers. In this work we demonstrate aba_b
%   the formation Y\_1 of a new type of polariton on the interface
%   between a cuprous oxide slab and a polystyrene micro-sphere placed

\begin{abstract}
Power dispatch is essential for providing stable, cost-effective, and eco-friendly electricity to society. However, traditional methods falter as power systems grow in scale and complexity, struggling with multitasking, swift problem-solving, and human-machine collaboration. This paper introduces GAIA, the pioneering Large Language Model (LLM) tailored for power dispatch tasks. We have developed a novel dataset construction technique that harnesses a range of data sources to fine-tune GAIA for optimal performance in this domain. This approach streamlines LLM training, allowing for the seamless integration of multidimensional data in power system management. Additionally, we have crafted specialized prompt strategies to boost GAIA's input-output efficiency in dispatch scenarios. When evaluated on the ElecBench benchmark, GAIA surpasses the baseline model LLaMA2 on multiple metrics. In practical applications, GAIA has demonstrated its ability to enhance decision-making processes, improve operational efficiency, and facilitate better human-machine interactions in power dispatch operations. This paper expands the application of LLMs to power dispatch and validates their practical utility, paving the way for future innovations in this field.
\end{abstract}

% \includegraphics{figs/grabs.pdf}

% \begin{highlights}
% \item Research highlights item 1
% \item Research highlights item 2
% \item Research highlights item 3
% \end{highlights}

% Keywords

\maketitle

\section{Introduction}
Ensuring power system stability and economic efficiency hinges on safe and effective power dispatch~\citep{valinejad2023computational}. System operators must skillfully balance generating unit outputs and load distribution across transmission lines, adapting to dynamic power supply and demand shifts due to human activities, weather variations, and emergencies. Dispatch processes must account for demand changes, generation unit costs and conditions, and transmission line capacities, all while meeting the power market's economic and reliability standards. The growing integration of renewable energy resources further complicates dispatch operations, demanding more advanced methods.

Traditional optimization algorithms~\citep{wood2013power} are effective for certain power dispatch problems but struggle with complexity, nonlinearity, and uncertainty. They also rely on precise model parameters, which are hard to maintain in dynamic power grids. In contrast, deep learning~\citep{wen2019optimal} and reinforcement learning~\citep{glavic2017reinforcement, cao2024survey} excel in processing high-dimensional data and adapting to changes in power grids. Yet, these methods fall short in generalization, frequently requiring retraining for new scenarios. Moreover, their ability to interact using natural language and learn from dispatchers' experiences through human-machine interaction is limited.

Recent breakthroughs in Large Language Models (LLMs) ~\citep{zhao2023survey} have revolutionized their capabilities. Models like Transformer, LLaMA~\citep{touvron2023llama}, and ChatGPT~\citep{chatgpt} have achieved mastery over language's deep structures and context through extensive pre-training, allowing them to comprehend and follow complex instructions. These LLMs can perform specific tasks at or above human levels when finely tuned and given detailed instructions. Prompt engineering~\citep{liu2023pre} further enhances their adaptability, enabling them to tackle new tasks without the need for large-scale retraining. This flexibility significantly benefits complex decision-making and human-machine interaction. LLMs' robust natural language processing skills are essential for efficient collaboration between humans and machines. Their strong generalization ability also minimizes the necessity for new models for different scenarios, showcasing their versatility and potential for widespread application.

Despite the success of LLMs in different domains, there has yet to be a dedicated LLM for power dispatch. Existing general-purpose LLMs like GPT-4~\citep{openai2023gpt4} or fine-tuned LLMs for the mathematics field like WizardMath~\citep{luo2023wizardmath} cannot adequately address the problems in power dispatch. The main barriers to building power dispatch LLM are listed as follows:
\begin{enumerate}
    \item \textbf{Lack of domain dataset}: The data in power systems is multiple perspective~\citep{ma2014deep}, such as load, cost, topology, etc. LLMs need to integrate these data to understand and learn the characteristics of power systems. The training data for other LLMs mostly consists of textual or basic numerical reasoning data, lacking actual power grid operation information.
    \item \textbf{Specific domain adaption}: LLMs for power dispatch need to be tailored to grasp the power sector's unique jargon and decision-making processes. This requires not only proficiency in specific terminology but also a comprehensive understanding of industry-specific scenarios, such as load forecasting and unit commitment, which are not inherent in general-purpose LLMs. 
    \item \textbf{Complex human-machine interaction}: The prompts for LLM in power dispatch scenarios significantly influence the performance of results. The prompts should be closely integrated with the power dispatch operations, containing key dispatching terminologies and conforming to related operational rules and decision-making processes. So that dispatchers can interact with LLM naturally using natural language and understand and trust LLM's results. 
    
\end{enumerate}

To overcome the above challenges, this paper seeks to enhance intelligent power dispatch and human-machine collaboration by advancing the use of LLMs in the energy sector. Our key contributions are:
\begin{enumerate}
    \item We develop GAIA, a model that uniquely fuses simulation numerical data with textual data, creating a dataset that encompasses operational and knowledge-based information, thus improving the model's applicability.
    \item We introduce a dedicated pipeline for data generation, processing, and LLM training tailored to power dispatch challenges. 
    \item We categorize power dispatch scenarios and design specific prompts for targeted training, incorporating techniques for interacting with LLMs, such as vector data representation, text data enrichment, and specialized terminology handling.
\end{enumerate}

Our evaluation of GAIA in the ElecBench benchmark~\citep{zhouelecbench} demonstrates its superiority over baseline model LLaMA2~\citep{touvron2023llama2}, across various dispatch-related metrics. The remaining sections of this paper are as follows: \Cref{Related Works} introduces the related works of domain-specific LLMs. The pipeline for GAIA is proposed in \Cref{Methodology}, including the division of dispatch scenarios, data generation and collection methods, and training techniques. \Cref{Performance} presents the model validation results, followed by \Cref{Conclusion}, which is the discussion and conclusion.

\section{Domain-specific Large Language Models}\label{Related Works}
Domain-specific LLMs have achieved significant progress in recent years. These advancements are largely attributed to the development of innovative dataset generation methods, efficient parameter fine-tuning techniques, and model customization for specific domains. The following contents summarize representative research in these key areas, collectively contributing to developing domain-specific LLMs.

\subsection{Specific Large Language Models}
Domain-specific LLMs have shown impressive capabilities in fields like mathematics, computer science, healthcare, and power systems. These models are refined from general LLMs using domain-specific data to meet particular needs.

For example, in the field of mathematics, WizardMath~\citep{luo2023wizardmath} is fine-tuned from the LLaMA2 model to enhance the mathematical reasoning ability, by generating diverse mathematical instruction data through Evol-Instruct. In computer programming tasks, CodeLLaMA~\citep{roziere2023code} uses the Self-Instruct method to generate datasets in the LLaMA2 70b model, which is then fine-tuned after validation. In the medical field, Med-PaLM~\citep{singhal2023large} utilized Instruction Tuning~\citep{zhang2023instruction} to fine-tune the Flan-PaLM~\citep{chung2022scaling} model efficiently, and its training data is obtained through random selection and manual evaluation filtering. For short-term load forecasting tasks in power systems, LFLLM~\citep{liulfllm} presents an efficient training method based on PEFT to address the challenging training problem of LLMs with massive parameters, ensuring excellent learning capability of the model.

The effectiveness of domain-specific LLMs stems from their strategic use and creation of enhanced datasets, along with advanced fine-tuning techniques, allowing them to tackle complex domain-specific reasoning tasks. Despite their proven potential across various sectors, there's a gap in the literature concerning data. While these models are adapted to particular applications using specialized domain datasets, there's a need for more systematic approaches in dataset generation, selection, optimization, and fine-tuning methods.

\subsection{Dataset Generation Methods}

Constructing a domain-specific LLM requires a series of steps to produce a relevant dataset. This involves selecting training tasks, analyzing and generating data, refining the dataset, and thorough filtering and post-processing to improve quality and diversity. The initial step is identifying the necessary training tasks for the domain. For instance, DARWIN~\citep{xie2023darwin}, a natural science LLM, categorizes its training tasks into material inverse design, property classification, and attribute regression prediction. During the data collection phase, domain-related papers, books, and news are typically converted into a processable text format. Subsequently, in the auto-generation of fine-tuning datasets phase, LLMs are often utilized to generate question-answer pairs. Among these, Self-Instruct~\citep{wang2023selfinstruct} enhances the model's ability to understand instructions by generating instructions by itself. Evol-Instruct~\citep{xu2023wizardlm} uses evolutionary algorithms to create complex code instructions, improving the model's fine-tuning performance on code generation tasks. Explore-Instruct~\citep{wan2023explore} employs an active learning strategy to explore and expand the specific domain's set of instructions, optimizing the model's task-processing performance. In the filtering and post-processing stage, irrelevant or low-quality data from the generated results are typically removed.

Existing research on domain-specific LLMs covers task selection, data collection, dataset generation, and post-processing. However, it frequently overlooks the unique needs of intricate fields like power system dispatch. These include handling specialized terms, complex formulas, and simulation data. Additionally, current methods often neglect to develop complex problem-solving skills from simulation data in dataset generation, a critical aspect for applications in power system dispatch.

\subsection{Baseline LLM}\label{Related Works Baseline}

Many open-source LLMs are available, each with unique strengths. MPT~\citep{MosaicML2023Introducing} excels in reasoning over long texts and offers optimized training and inference speeds. Falcon~\citep{almazrouei2023falcon} enhances performance by selectively processing network data for pre-training datasets. Meta's LLaMA2, an improved version of LLaMA, demonstrates strong logical and mathematical reasoning capabilities. In benchmarks like MMLU~\citep{hendrycks2021measuring} and MATH~\citep{hendrycks2021measuringmath}, LLaMA2 outperforms competitors with similar model sizes.

LLaMA2 is an LLM based on the Transformer Decoder architecture, which has been optimized on top of the standard structure, such as introducing RMSNorm~\citep{zhang2019root} pre-normalization layer, SwiGLU~\citep{shazeer2020glu} activation function, and Rotary Position Embedding (RoPE)~\citep{su2023roformer}. The training dataset for LLaMA2 is massive, reaching 20 trillion tokens, which is a 40\% increase compared to the previous generation model, and the context length has been extended from 2048 to 4096 tokens. This allows the model to understand and generate longer texts.

A notable feature of the LLaMA2 is the adoption of the Grouped-Query Attention ~\citep{ainslie2023gqa} mechanism. This optimized attention mechanism improves inference throughput by reducing the size of the key-value cache. The Grouped-Query Attention reduces memory usage by sharing a single key-value projection, and experiments have shown that it performs comparably to Multi-Head Attention ~\citep{cordonnier2020multi} on most evaluation tasks and typically outperforms Multi-Query Attention ~\citep{shazeer2019fast}.

To enhance the model's performance, LLaMA2 also uses the Reinforcement Learning from Human Feedback ~\citep{ouyang2022training} method, which iteratively fine-tunes the model with human feedback data, including using rejection sampling and Proximal Policy Optimization ~\citep{schulman2017proximal}.

In summary, the LLaMA2 excels in various benchmark tests due to its sophisticated architecture and training techniques. Its versatility and precision in generating responses establish it as a significant benchmark in the LLM field.

\subsection{Parameter Efficient Fine-tuning Methods}
The Parameter-Efficient Fine-Tuning (PEFT) ~\citep{xu2023parameterefficient} is a class of fine-tuning methods for computational resource-limited situations. Selecting the suitable PEFT method is crucial for improving the LLM performance of downstream tasks. Adapter~\citep{houlsby2019parameterefficient} achieves parameter-efficient task-specific adjustments by inserting small network layers into pre-trained models. Still, this method may introduce additional computational overhead and may be less effective than full model fine-tuning in some tasks. P-Tuning~\citep{liu2023gpt} fine-tunes models by adding learnable continuous vectors (i.e., prompts). Although this method has advantages in parameter efficiency, it relies on the model's sensitivity to prompts and may require larger training datasets to optimize prompts.

To address these limitations, Low-Rank Adaptation (LoRA) ~\citep{hu2021lora} reduces the number of fine-tuning parameters and alleviates computational burden by performing low-rank decomposition on weight matrices and updating small matrices. Further, Quantized Low-Rank Adaptation (QLoRA) ~\citep{dettmers2023qlora} decreases the representation precision of low-rank matrices using quantization techniques, which reduces not only the number of parameters but also the model's memory footprint and computational complexity, making the fine-tuning process more efficient in resource-constrained environments. These methods demonstrate the ability to adapt to specific tasks and to preserve pre-trained model knowledge by fine-tuning a few parameters, highlighting the importance of choosing suitable fine-tuning strategies in different scenarios.

\section{Methodology}\label{Methodology}
This section outlines a robust methodology for creating an LLM tailored to power dispatch tasks. Our goal is to build a model that comprehends power system dynamics, facilitates human-computer interaction, and delivers precise decisions. We employ a multi-stage pipeline—comprising data generation, preprocessing, training, and optimization leveraging multi-task learning and targeted training to refine the LLM's capabilities in the power sector. We detail the training data composition and methods for processing text and simulation data, offering theoretical and practical foundations for the model's development.

\begin{figure*}[!h]
\centering
\includegraphics[width=1\textwidth]{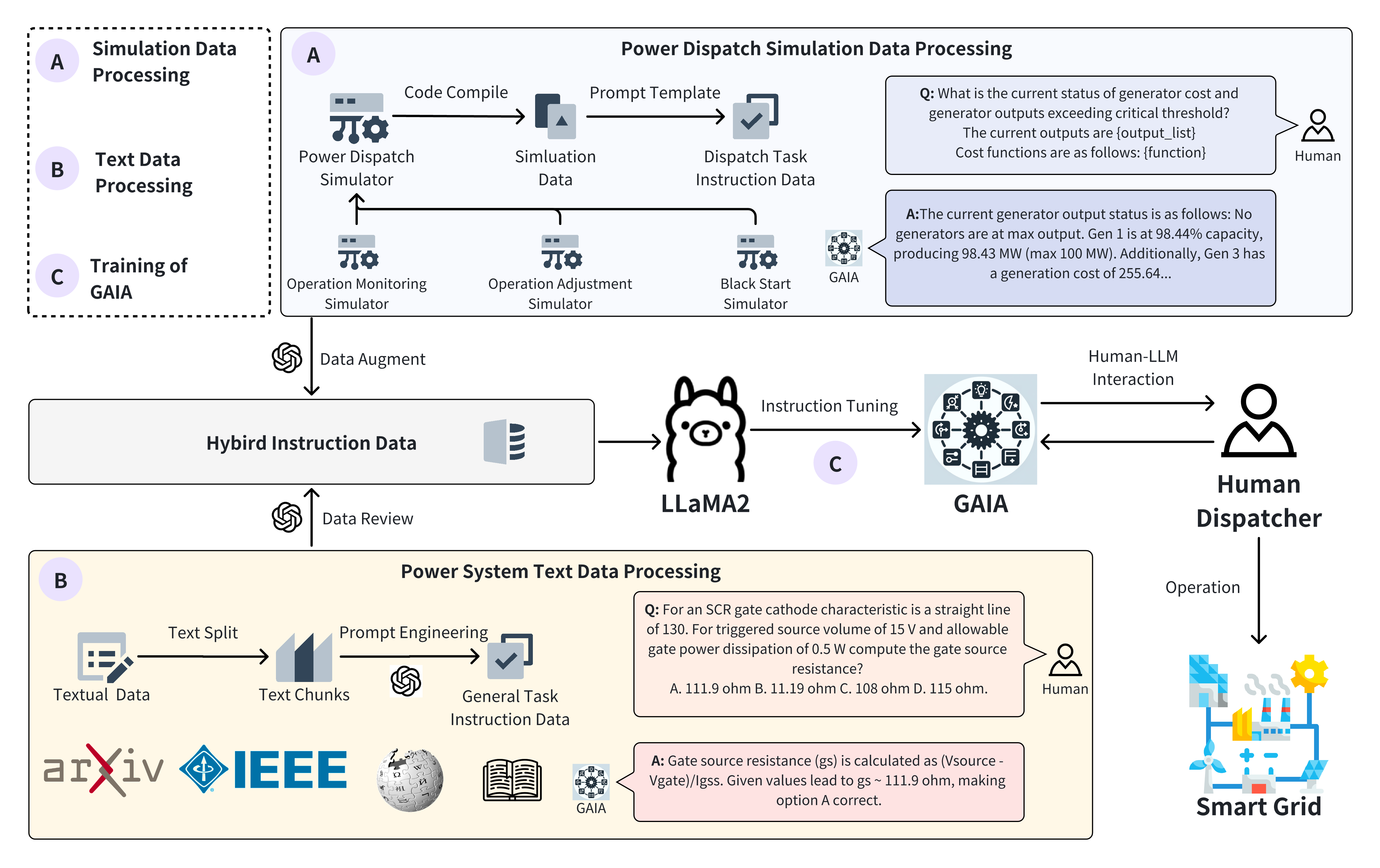}
\caption{Overall Framework of GAIA} \label{fig_framework}
\end{figure*}

\subsection{Framework}

The objective of building an LLM for power dispatch is clear: the LLM can grasp power system dynamics, aid dispatchers through human-machine collaboration, and consistently make precise decisions. Current LLMs struggle with the power industry's unique challenges, such as interpreting diverse power system data, specialized jargon, and real-world dispatch scenarios.

Addressing these issues, we hypothesize that an LLM, trained with a robust power dispatch dataset through multi-task learning and specialized fine-tuning, will enhance its predictive capabilities and decision support for dispatchers. We've devised a multi-stage pipeline: data generation and preprocessing, interactive prompt design, and model training and optimization. This approach seeks to surpass traditional power dispatch limitations and bolster intelligent power system operations.

As shown in \Cref{fig_framework}, the overall framework's pipeline includes the following three key steps:
\begin{enumerate}
    \item \textbf{Simulation Data Processing}: 
    We select simulation scenarios that align with dispatch operations: adjustment, monitoring, and black start. Each scenario has a dedicated simulation program to produce system data. This data is then transformed into structured Q\&A pairs using templates, enhancing the model's comprehension and learning.
    \item \textbf{Text Data Processing}: 
    We employ Optical Character Recognition (OCR) to digitize power-related documents, converting them to editable text. This text is then segmented and paired with prompts. Finally, GPT-4 generates additional Q\&A pairs, enriching the training dataset.
    \item \textbf{Training of GAIA}: 
    The LLaMA2 undergoes supervised fine-tuning with extensive Q\&A data from simulations and texts. This targeted training sharpens the model's performance on power dispatch tasks, ensuring greater efficiency and precision in real-world operations.
\end{enumerate}

\subsection{Data Collection and Preprocessing}
This pipeline combines text data and simulation numerical data generation of power systems, thereby not only improving the model's understanding of power system data but also enhancing its generalization ability and human-machine interaction efficiency in practical applications.

\subsubsection{Simulation Data Processing}
\begin{figure}[h]
\centering
\includegraphics[width=0.5\textwidth]{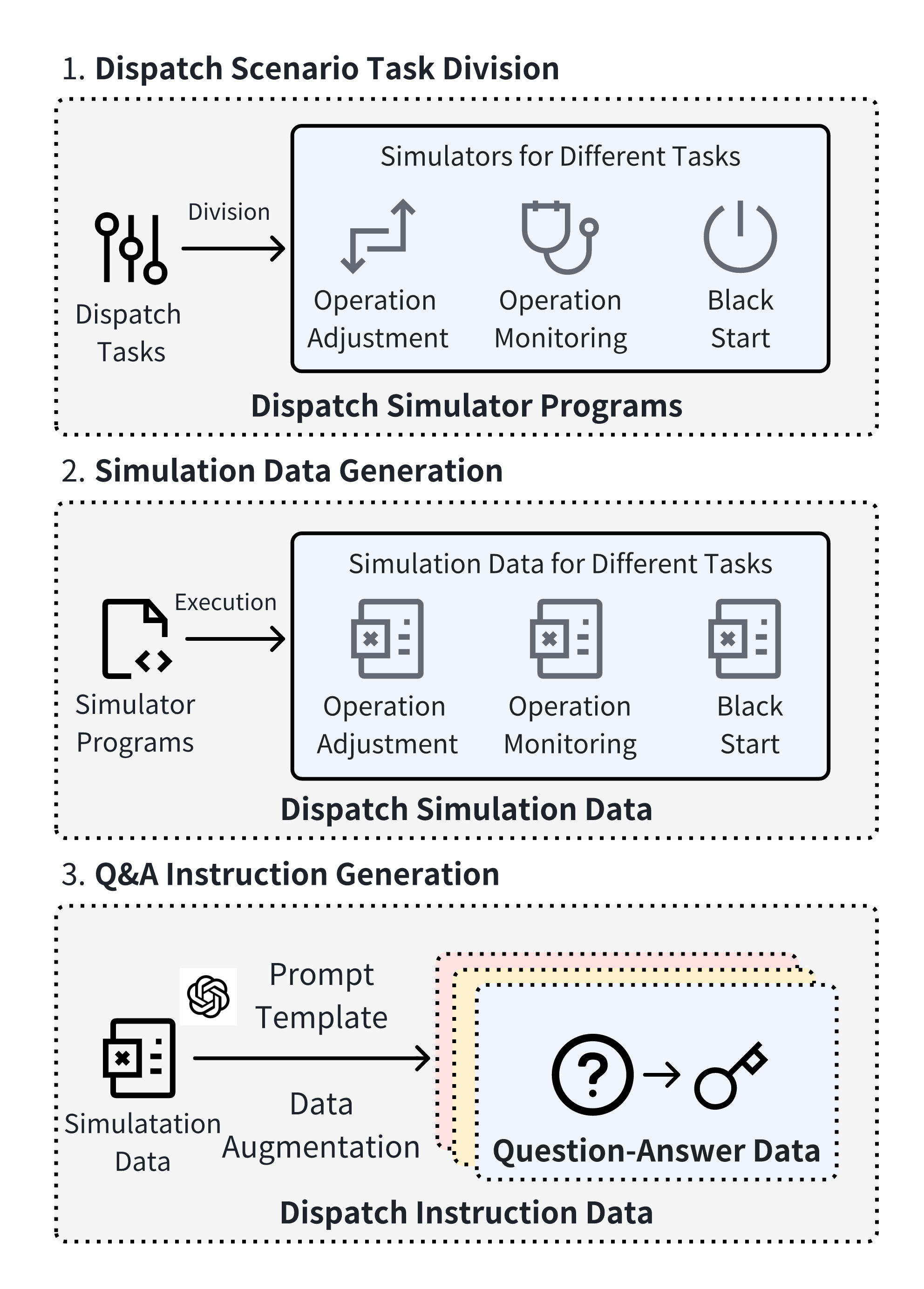}
\caption{Simulation Data Processing Flow} \label{fig_simulator}
\end{figure}

Simulation data is essential for developing an LLM for power dispatch, enabling it to grasp the physical dynamics of power systems. This data is predominantly numerical and must be converted into actionable Q\&A training material. The conversion entails creating realistic power dispatch scenarios across different contexts and turning the numerical simulations into text data for training.

The process encounters three primary challenges. The first is accurately capturing the complexity and dynamics of power dispatch in the simulation data. The second challenge is producing high-quality simulation data that meets real-world operational standards. Lastly, crafting effective Q\&A pairs that mimic the interaction between dispatchers and the power system is challenging.

A new method for processing simulation data is proposed to address the aforementioned issues. This method is divided into three steps: dispatch scenario task division, simulation data generation, and Q\&A instruction generation. The processing flow is shown in \Cref{fig_simulator}.

\paragraph{Dispatch Scenario Task Division}

First, in the dispatch scenario task division stage, the dispatch scenarios are divided into operation, diagnosis, and recovery based on the actual operation requirement of power system dispatch and their corresponding mathematical problem forms~\citep{lee1985united, gungor2012survey}. Due to the limitation in computational resources, operation adjustment~\citep{chowdhury1990review}, operation monitoring~\citep{gao2015survey, bevrani2014power}, and black start~\citep{patsakis2018optimal} are chosen according to the subdivided tasks frequency in dispatch operation. Specifically, economic dispatch focuses on minimizing generation costs while meeting load demand and its mathematical expression is an optimization problem, which linear or nonlinear programming methods can solve; operation monitoring focuses on ensuring the stability and reliability of system operation, and the mathematical problems involve probability-based risk assessments; black start focuses on the rapid and safe recovery of the power system after a complete crash, the mathematical problems to be solved include sequential decision-making and path planning, etc.

\paragraph{Simulation Data Generation}

During the simulation data generation stage, the specialized simulator programs are established for each divided task. These programs are designed to simulate system conditions under different scales of power systems and load disturbances. Using the Monte Carlo sampling method~\citep{hastings1970monte}, a large volume of load data is generated under various operating conditions to simulate the real-world load uncertainties. This method allows us to examine the various possibilities for power system dispatch and guarantee the system's robustness under different load levels and system states. A detailed description of the simulation data generation methods for each specific scenario is as follows:
\begin{itemize}
    \item \textbf{Operation Adjustment}: 
    This task aims to balance system safety and operational economy by adjusting generator outputs when safety limits are breached. The process involves modeling node-specific loads and renewable energy distributions for various system sizes, using Monte Carlo simulations to mimic real-world variability. The Alternating Current Optimal Power Flow (AC-OPF) is then computed with this data to set power outputs and flows. To replicate operational events, the capacity of the top three loaded lines is randomly reduced.
    The goal here is to forecast operation outcomes and flag potentially harmful actions. We simulate different power system scales by varying node loads within a defined range. The AC-OPF, based on Monte Carlo sampled data, assesses system conditions. A high-flow line is then randomly disconnected to model dispatcher actions. Safety is confirmed if the AC-OPF converges post-disconnection; divergence indicates an unsafe operation.

    \item \textbf{Black Start}: 
    The objective is to establish a generator and node restoration sequence following a total blackout. The simulation scales to different system sizes and employs a Genetic Algorithm to sequence generator bus recovery. The Single Source Shortest Path algorithm sequences node restoration. The final step is integrating these sequences to determine the optimal recovery order for all generators and nodes.
\end{itemize}

These simulator programs can generate simulation data that matches actual power dispatch tasks. These data not only include various operating conditions and dispatch parameters but also simulate various events that may occur during operation.

\paragraph{Q\&A Instruction Generation}
In the Q\&A instruction generation phase, we tailor instructions to match specific dispatch tasks, employing distinct question-and-answer templates. For instance:
\begin{itemize}
    \item \textbf{Operation adjustment tasks}: Instructions prompt the LLM to calculate each generator's generation costs and outputs.
    \item \textbf{Operation monitoring tasks}: Instructions direct the model to evaluate system safety boundaries and suggest operational adjustments.
    \item \textbf{Black start tasks}: Instructions guide the model in determining the startup sequence for generators and nodes.
\end{itemize}

Details on the instruction design are found in \Cref{Prompt_Instruction}. Moreover, we leverage GPT-4 for data augmentation, enhancing text descriptions to facilitate dispatcher interaction via natural language.

\vspace{1em}

Through these simulation data processing steps, high-quality and highly realistic training data are provided for LLM, laying a solid foundation for the model's effective training and subsequent practical applications.

\subsubsection{Text Data Processing}

\begin{figure}[h]
\centering
\includegraphics[width=0.5\textwidth]{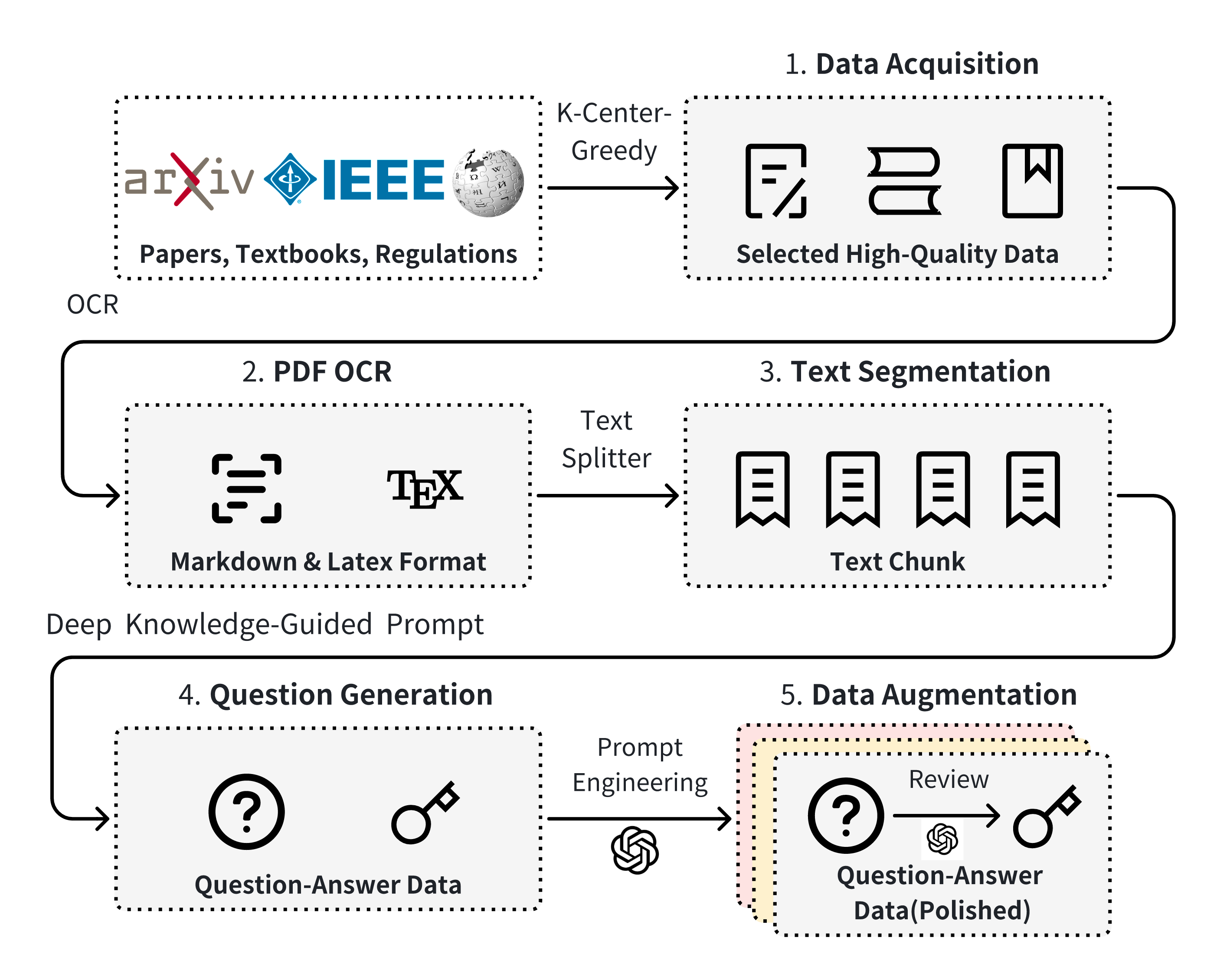}
\caption{Text Data Processing Flow} \label{fig_text}
\end{figure}

Building GAIA requires professional, comprehensive, and accurate text data. This entails extracting high-quality text from sources like papers and textbooks. Challenges include the prevalence of technical terms and complex formulas in power systems literature, the presence of irrelevant or redundant content, and the need for an efficient method to create Q\&A pairs for model training.

Current methods use automated tools to extract text and generate question-and-answer pairs through keyword matching. However, these techniques often fail to grasp the nuanced knowledge of power systems and struggle with complex formulas, leading to inconsistent training data quality.

To address these issues, a new text data processing flow is proposed and shown in \Cref{fig_text}. The whole process contains five steps: data acquisition, PDF OCR, text segmentation, question generation, and data augmentation. 

Through this series of steps, we have established a complete text-processing pipeline from literature screening to question generation. Relevant literature is carefully selected using specific algorithms, and computational formulas are retained during document parsing. Subsequently, through meticulous text segmentation and innovative question generation methods, as well as data augmentation and quality review, we ensure the richness and high quality of information, laying a solid foundation for the model's logical computing capabilities and general knowledge of power systems. The detailed steps are as follows:

\begin{enumerate}
    \item In the data acquisition stage, the K-Center-Greedy~\citep{sener2017active} algorithm is utilized to cluster literature topics and filter out high-quality literature that is highly relevant to power systems and has less information redundancy. This literature includes research papers, industry statutes, and authoritative textbooks, providing a solid theoretical foundation, operational procedures, and cutting-edge research results for the model.
    \item Then, PDF files are converted into editable markdown format in the data processing stage using OCR technology. We pay special attention to converting mathematical formulas in the literature into Latex format using the Nougat~\citep{blecher2023nougat} to enhance the LLM's mathematical reasoning ability. 
    \item Semantic-aware text segmentation ensures that each text block is divided into suitable text blocks based on content and semantics.
    \item In the question generation stage, Deep Knowledge-Guided Prompt, an improvement over the existing Self-Instruct method is utilized. Each segmented text block is input into GPT-4, and these prompts are applied to guide the model in generating questions and answers based on specific background knowledge performing computational reasoning when necessary. 
    \item In addition, data augmentation techniques are utilized to enrich the diversity of training data by transforming sentence structures and expanding question backgrounds. Finally, GPT-4 is applied as a Reviewer to polish the quality of generated questions and answers to ensure the accuracy of the data.
\end{enumerate}

Through this series of text data processing procedures, we have laid a solid foundation of background knowledge in power systems for training the GAIA and have improved the model's understanding and handling capabilities for power dispatch issues.

\subsection{Prompt Designing}\label{Prompt_Instruction}
The design of the Prompt is crucial to the understanding and generation ability of LLM during the instruction data generation stage in the GAIA training phase and the specific power scheduling problem-solving stage in the inference phase. The main issue currently faced is how to construct effective Prompts to facilitate LLM's accurate understanding of complex power system scheduling tasks and generate useful outputs.

The current practice is to have LLM work within a restricted context, for example, generating questions and answers around a specific text passage, or providing a simplified reasoning process when dealing with computational problems. This approach often fails to fully leverage the capabilities of LLMs, especially when dealing with tasks that involve a high level of expertise and complex logic.

\begin{figure}[h]
\centering
\includegraphics[width=0.5\textwidth]{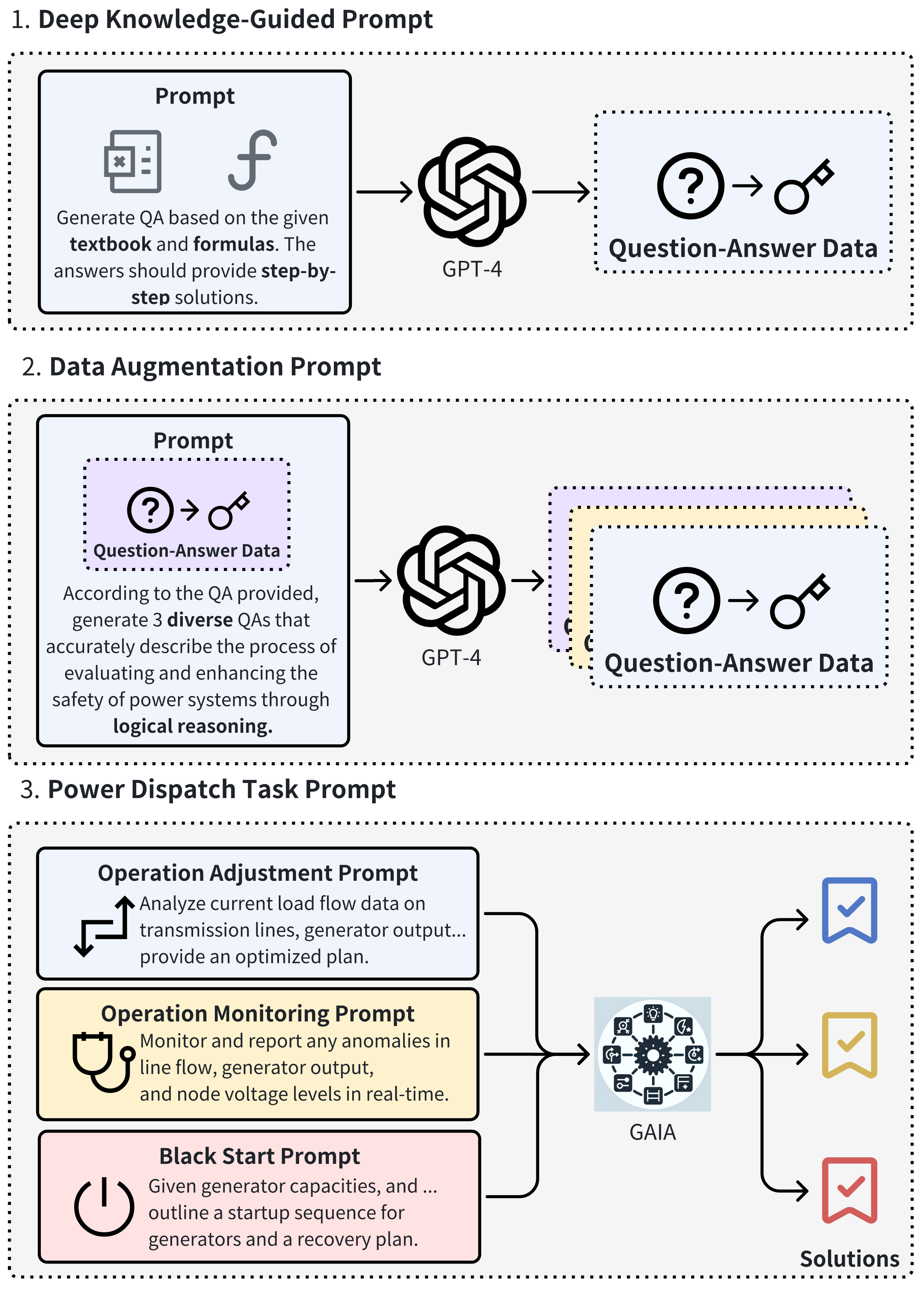}
\caption{Prompt Designing} \label{fig_prompt}
\end{figure}

To address this issue, we have adopted a new Prompt design method, as shown in \Cref{fig_prompt}, to improve the model's performance in power dispatch tasks. The following is a detailed description of the prompt construction methods designed for different tasks:

% The main issue for GAIA is how to construct effective prompts under limited context length to help the model accurately understand complex power system conditions and task objectives and how to design the differential prompts for different tasks.

\subsubsection{Deep Knowledge-Guided Prompt}
The Deep Knowledge-Guided Prompt is used during the GAIA training phase to create Q\&A instructions. GPT-4, acting as a power system expert, is limited to producing questions and answers related to the provided textbook paragraph and power system dispatch problem. For content with formulas or charts, we highlight this in the prompt and instruct GPT-4 to offer detailed, step-by-step reasoning for calculation questions~\citep{wei2023chainofthought}. This approach enhances the depth and logic of the questions GPT-4 generates.

\subsubsection{Data Augmentation Prompt}
In the design of data augmentation Prompts, the goal is to re-describe the same concept or problem in different ways and ensure the accuracy and logical rigour of the information. For this target, the GPT-4 is asked to generate diversified expressions and then check the data's logical reasoning and calculation process to ensure the safety requirements of the power system. This prompt can improve the GPT-4's performance in safety-critical tasks.

\subsubsection{Power Dispatch Task Prompt}
The most important issue for the prompt design of dispatch task prompts is to solve the problem of how to input the structured information. First, the category of the tasks is clarified. Then, the relevant background and data, especially numerical data, are provided and presented in a list format for better parsing by the model. In the prompt design for operation adjustment, operation monitoring, and black start, we not only combined the needs of dispatchers but also assumed that the model should think like power system experts.

\paragraph{Operation Adjustment}
The prompt design for operation adjustment should reflect the need for real-time monitoring and dynamic operation adjustment of power systems and aim to optimize the grid's operation by comprehensively considering the status of lines, generators, and nodes. For example, the LLM needs to monitor the load flow of transmission lines to promptly detect and handle overload or near-overload situations, monitor the output power of generator sets to ensure that generators are operated within cost-benefit and safety boundaries, and monitor the voltage conditions of power grid nodes to prevent voltage from exceeding the normal operating range. Furthermore, the generators' output adjustment plans are based on the actual load situation and line capacity limits to achieve grid load balance and ensure system stability.

\paragraph{Operation Monitoring}
The operation monitoring task prompt design mainly focuses on real-time status monitoring and anomaly detection of power grids. This includes monitoring the flow of transmission lines to identify overload risks, assessing the real-time output of generators to ensure safe operation, and monitoring node voltages to prevent voltage anomalies. In addition, the task involves evaluating whether the power system can be restored to a stable state by adjusting generator outputs when any anomaly is detected.

\paragraph{Black Start}
The prompt design for black start tasks aims to ensure the priority supply of critical loads and gradually achieve a robust recovery of the entire grid. First, the startup sequence of generator sets is determined considering each generator's capacity and power ramp rate, as well as their connections to load nodes, to ensure the stability and efficiency of the recovery process. Then, based on the generator startup sequence, the recovery sequence of nodes is further determined. The critical load nodes are restored first, considering line recovery time and additional nodes that are beneficial to system recovery. This prompt design takes into account the key factors of grid recovery.

Through such prompt designs, the LLM is guided to handle complex problems of power dispatch more effectively, improving the model's practicality and reliability.

\subsection{Training of GAIA}
In the training phase of GAIA, LLaMA2 is selected as the base model, and the LoRA method is employed for subsequent fine-tuning. With a carefully designed training dataset, GAIA can deeply understand the operating principles of power systems and effectively execute power scheduling tasks. In addition, we also explore different model parameter sizes and fine-tuning methods to learn the specific tasks of power scheduling.

\subsubsection{Baseline LLM}
In exploring LLMs suitable for the power dispatch problems, LLaMA2 is selected as the base model, as mentioned in \Cref{Related Works Baseline}, mainly due to its outstanding performance in multiple reasoning benchmark tests, and the advanced architecture and training methods. The LLaMA2 model adopts the latest architecture and is designed to capture and understand rich semantic information, which is crucial for comprehending professional terminology and complex data in the power industry. Furthermore, LLaMA2's training method focuses on multi-task learning, which helps the model better generalize and adapt to different data distributions when handling various power dispatch tasks. During the fine-tuning process, the LLaMA2 model can effectively utilize simulation data and knowledge texts from the power dispatch domain to gradually build a deep understanding of the power systems dynamics. This specialized training enables LLaMA2 to generate more accurate and reasonable responses for specific tasks such as power dispatch, thus providing effective decision support in practical applications.

\subsubsection{Composition of Training Data}
\begin{figure}[h]
\centering
\includegraphics[width=0.5\textwidth]{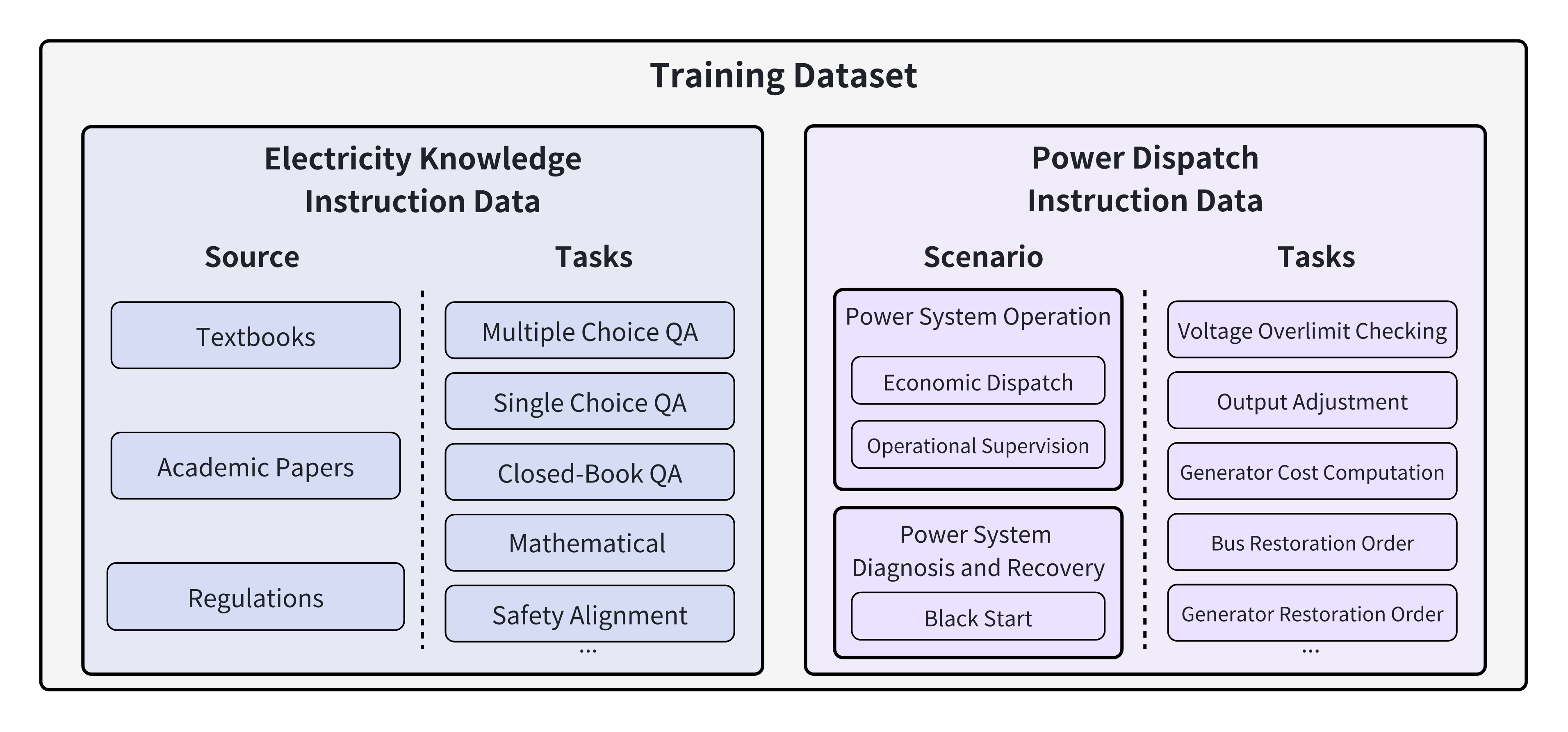}
\caption{Training Dataset Composition} \label{fig_training_data}
\end{figure}
In constructing the GAIA, the training dataset is carefully designed to ensure that the model can comprehensively understand the operation principles of power systems and effectively perform power dispatch tasks. As shown in \Cref{fig_training_data}, The training dataset contains a total of 160,000 fine-tuning data, consisting of two main parts: power knowledge and power dispatch data. Power knowledge data accounts for 15\% and power dispatch data accounts for 85\%.

The power system knowledge data comes from textbooks, academic papers, and industry regulations in the power system domain. These data are sourced from high-quality online resources such as Wikipedia, arXiv, and IEEE. These tasks include multiple-choice, single-choice, and question-and-answer tasks on basic concepts and mathematical reasoning problems. The average input length for each data entry is approximately 40 tokens, and the output length is 234 tokens. In addition, to ensure that the LLM follows the values and standards of the power industry when dealing with safety-related issues, we specifically designed value alignment tasks related to safety issues. These tasks aim to improve the LLM's generalization and reasoning capabilities, enabling it to handle more complex problems based on understanding the basic knowledge of power systems.

Power dispatch data is generated based on specific scenarios selected from two dispatching business scenarios: Power System Operation and Power System Diagnosis and Recovery. Operation Adjustment and Operation Monitoring are chosen for power system operations. For Power System Diagnosis and Recovery, Black Start is chosen. The specialized simulation programs for power systems of different scales, including IEEE 14-bus system, IEEE 30-bus system, IEEE 57-bus system, and IEEE 118-bus system, are built for each scenario. These programs can simulate the behavior of power systems under various operating conditions. Through these simulation programs, we have generated a large amount of numerical data and formed various training tasks based on this data. The average input length for each data entry is approximately 476 tokens, and the output length is 167 tokens. These tasks include voltage boundary issues, output adjustments, generation cost calculations, and the start-up sequence of generators and buses. The design of these tasks aims to enable the LLM to make accurate judgments and effective decisions for specific operating situations during actual dispatching processes.

Overall, the composition of the training data considers both the depth and breadth of power knowledge and the complexity of actual power-dispatching operations. With such a combination of training data, we expect the GAIA to effectively complete various power dispatch tasks based on understanding power systems.

\subsubsection{Fine-tuning Method}

% Three GAIA models with different parameter sizes are fine-tuned during the fine-tuning stage: 7b, 13b, and 70b. Also, a comprehensive evaluation of various efficient parameter fine-tuning methods, including P-Tuning, Adapter, LoRA, and QLoRA, is conducted. After comparison, LoRA is chosen for fine-tuning. The core advantage of LoRA is that it introduces low-rank matrices to modify the weights of pre-trained models, which not only significantly reduces the number of parameters that need to be trained but also maintains the flexibility and expressiveness of the model. This is particularly important for specialized fields such as power dispatch, as it requires the model to understand and process highly specialized knowledge and data while also demanding the model to have a faster ability to adapt to new tasks.
During the fine-tuning stage, we fine-tune three GAIA models with varying parameter sizes: 7b, 13b, and 70b. We also rigorously evaluate several efficient fine-tuning methods, including P-Tuning, Adapter, LoRA, and QLoRA. LoRA emerges as the preferred method due to its strategic use of low-rank matrices that adjust the weights of pre-trained models. This approach drastically cuts the number of trainable parameters while preserving the model's flexibility and expressiveness. Such efficiency is crucial in specialized domains like power dispatch, where the model must grasp intricate knowledge and data and swiftly adapt to new tasks.

\section{Performance Analysis for Power-Related Tasks}\label{Performance}
In this section, GAIA's performance in power dispatching was evaluated through the ElecBench evaluation system. Although GPT-4 demonstrates excellent performance and has become an important benchmark, we mainly focus on the comparison with the basic models, considering the professionalism and data privacy issues in practical applications. The results show that GAIA-70b outperforms LLaMA2 and GPT-3.5 in key performance indicators, especially in terms of accuracy and stability in operation monitoring and power dispatching. The case analysis emphasizes its professional adaptability in power system analysis. In addition, the LoRA fine-tuning method shows outstanding performance in training efficiency and performance convergence. These findings highlight the potential of GAIA as a professional tool for power dispatching.

\subsection{Metric}

We evaluated GAIA's power dispatch performance using the ElecBench system, which assesses large models across six key areas: factuality, logicality, stability, fairness, safety, and expressiveness. These areas are further broken down into twenty-four sub-dimensions tailored to specific power dispatch tasks. ElecBench offers a thorough and nuanced analysis, capturing both the practical value and potential issues of the model within the power system sector.

ElecBench employs both objective and subjective test questions, including judgment, multiple-choice, and Q\&A tasks. GPT-4 serves as an impartial scorer, enhancing the evaluation's efficiency and objectivity. This objective approach swiftly yields performance metrics for GAIA, particularly in factuality and logicality.

\subsection{ElecBench Evaluation Outcomes}

Under the ElecBench evaluation framework, a comprehensive performance assessment of GAIA with three different parameter sizes is conducted and compared with several other well-known LLMs, including LLaMA2, GPT-3.5, and GPT-4. Through a series of standard tests and case analyses, we found that GAIA-70b outperforms LLaMA2 and GPT-3.5 in most indicators in power dispatch, slightly below GPT4.
\input{score_table}

\subsection{Case Study}

\begin{figure}[h!]
\centering
\includegraphics[width=0.5\textwidth]{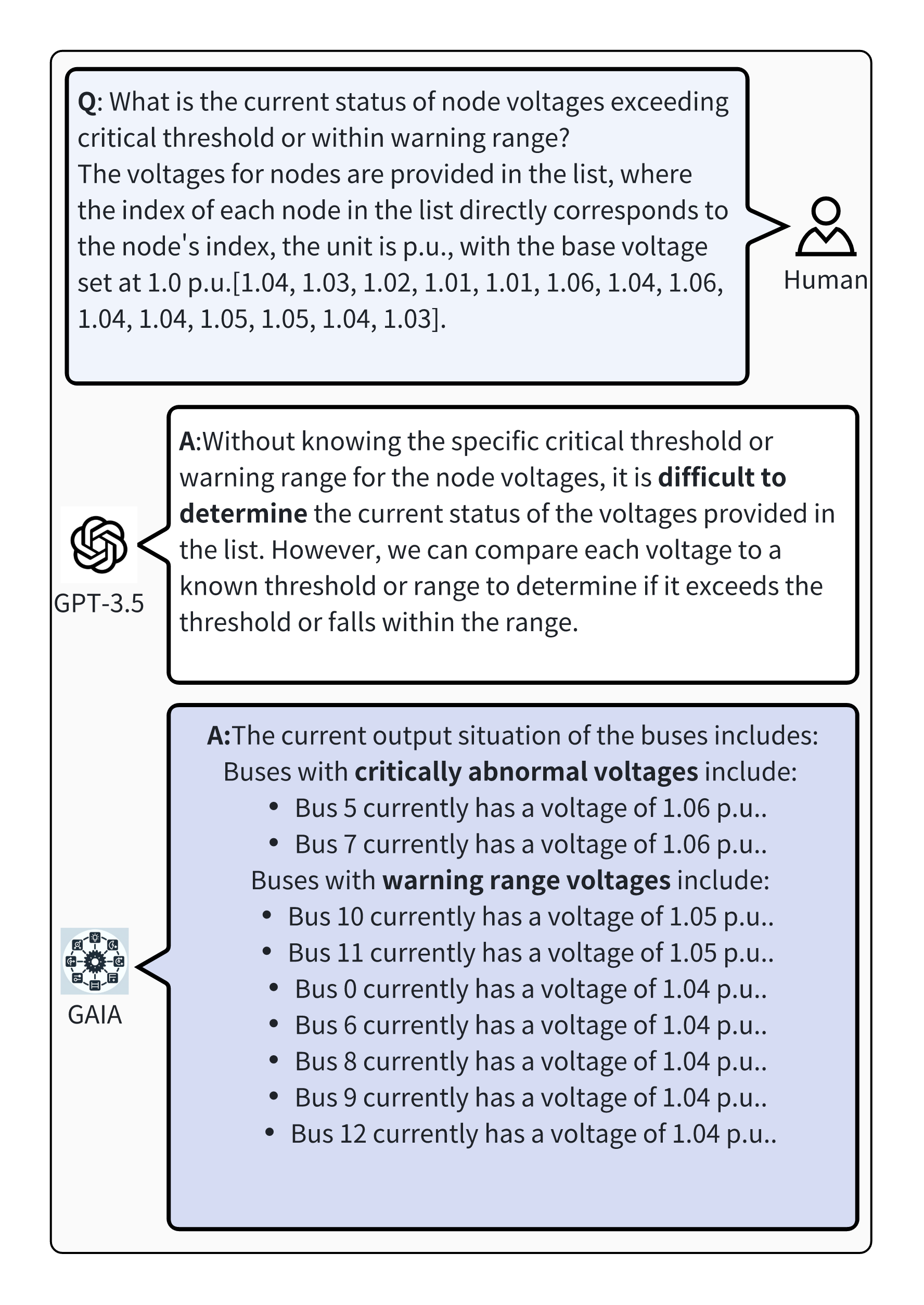}
\caption{Case Study of GAIA in Dispatch Task} \label{fig_case_study}
\end{figure}

In the case study, illustrated in \Cref{fig_case_study}, we compared GAIA-70b with GPT-3.5 in addressing voltage boundary issues during power dispatch. GAIA demonstrated superior precision in pinpointing nodes with abnormal voltage levels, while GPT-3.5 offered a broader, less detailed analysis. Unlike GPT-3.5, GAIA explicitly flags nodes with critical voltage values and those within warning thresholds. This precision suggests that GAIA incorporates optimized algorithms or industry-specific standards tailored for power system analysis. Such targeted diagnostics are vital for power system operators, enabling swift identification and correction of potential instabilities or failures.

\subsection{Fine-tuning Methods Evaluation}

\begin{figure}[h]
\centering
\includegraphics[width=0.45\textwidth]{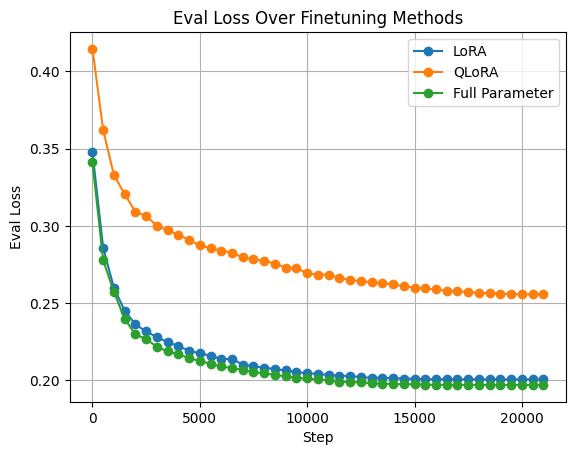}
\caption{Eval Loss Comparison on Different Finetuning Methods} \label{fig_finetune}
\end{figure}

When evaluating different fine-tuning methods, we compared the effects of Lora, QLoRA, and full-parameter fine-tuning, as shown in \Cref{fig_finetune}. We validated various fine-tuning methods using Loss by testing the 7b parameter scale model. The results show that although QLoRA provides a faster training speed, it is not as good as LoRA and full-parameter fine-tuning in terms of Loss convergence. Full-parameter fine-tuning offers better performance improvements but takes a longer training time. LoRA offers an accelerated training process while simultaneously achieving satisfactory loss convergence.

\section{Conclusion}\label{Conclusion}

This paper introduces GAIA, the first Large Language Model (LLM) tailored for the power dispatch sector, and evaluates its performance using the ElecBench framework. It highlights key performance influencers such as data quality, prompt design, and fine-tuning methods. Optimizing these aspects is essential for maximizing GAIA's practical benefits in power dispatch. The model showcases substantial potential in real-world scenarios, adeptly managing complex dispatch tasks and offering valuable insights into economic and stability considerations. This underscores the pivotal role of advanced AI technologies in digitally transforming traditional industries, enhancing operational efficiency, and ensuring the reliability of power systems.

The model currently faces limitations, including the need for enhanced language and logic capabilities, as well as improved robustness in extreme conditions and rare events. Future efforts will focus on expanding the model's industry knowledge, refining fine-tuning techniques, and boosting its practical efficiency and accuracy, all while prioritizing safety. By fostering technological innovation and interdisciplinary collaboration, we aim for smarter and more sustainable advancements in the power dispatch.

\bibliographystyle{cas-model2-names}
% \bibliographystyle{cas-model2-names}
% \bibliographystyle{elsarticle-num}
\bibliography{references}

\input{appendix}
\end{document}

%% file: score_table.tex
\subsubsection{Evaluation on Operation Monitoring}

\begin{table}[h!]
\centering
\caption{Evaluation on Operation Monitoring}
\small
\label{table_monitoring}
\begin{tabular}{lcccc}
\toprule
 & \textbf{Factuality} & \textbf{Logicality} & \textbf{Safety} & \textbf{Stability} \\
\midrule
GPT-4 & \textbf{8.333} & \textbf{8.920} & 9.000 & \textbf{8.860} \\
GPT-3.5 & 7.351 & 8.040 & 8.963 & 7.820 \\
LLaMA2-70b & 6.875 & 7.580 & 9.519 & 7.780 \\
LLaMA2-13b & 6.891 & 7.260 & 9.565 & 7.460 \\
LLaMA2-7b & 6.466 & 6.680 & 9.227 & 6.440 \\
\textbf{GAIA-70b} & 7.704 & 7.940 & \textbf{9.806} & 8.060 \\
\textbf{GAIA-13b} & 8.091 & 7.260 & \textbf{9.806} & 6.880 \\
\textbf{GAIA-7b} & 7.671 & 7.320 & 9.764 & 6.540 \\
\bottomrule
\end{tabular}
\end{table}

\begin{figure}[h!]
\centering
\includegraphics[width=0.48\textwidth,height=0.48\textwidth]{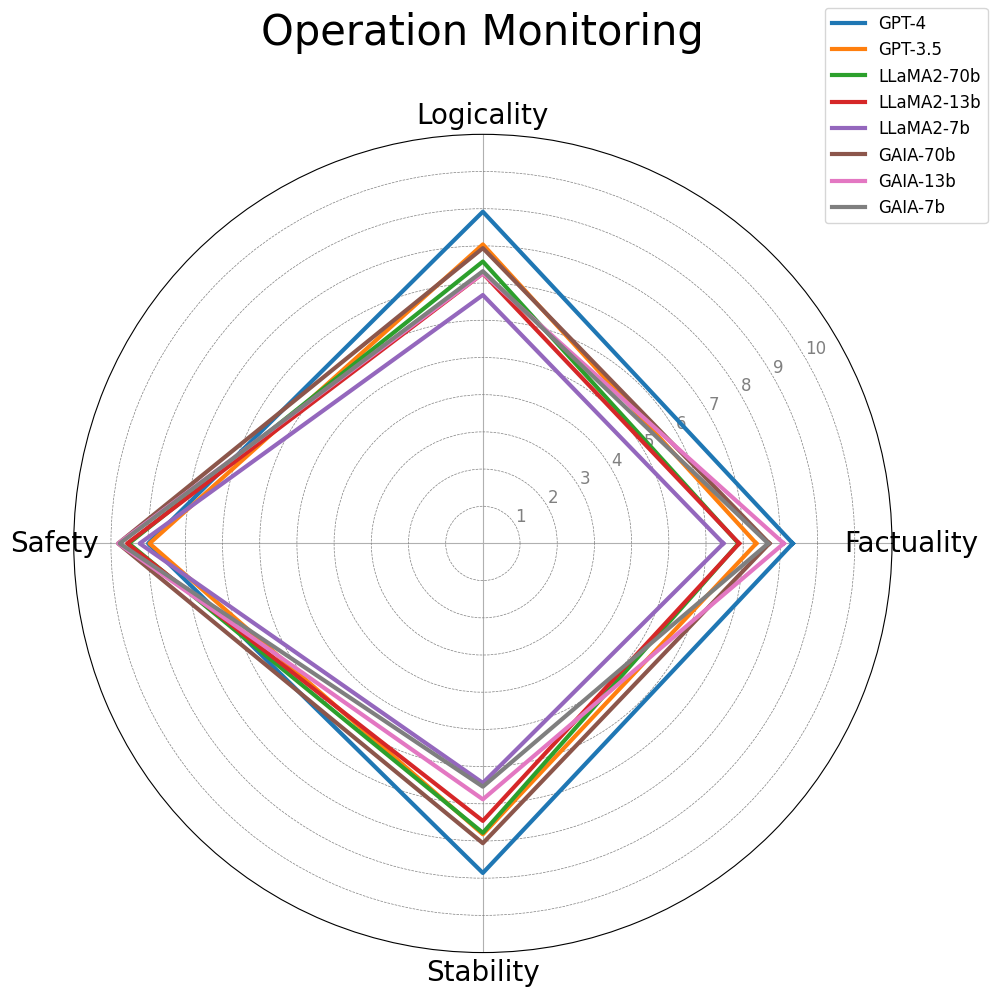}
\caption{Evaluation Result on Operation Monitoring} \label{fig_eval_operationmonitoring}
\end{figure}

% As shown in \Cref{table_monitoring} and \Cref{fig_eval_operationmonitoring}, in the realm of operation monitoring, the GAIA-70b model showcases remarkable performance across various metrics, as evidenced in recent evaluations. Notably, GAIA-70b excels in the category of safety, achieving the highest score of 9.806 among the models tested. This underscores its superior ability to generate reliable outputs and minimize potential risks, making it an exemplary choice for applications where safety is paramount. While its scores in factuality (7.704), logicality (7.940), and stability (8.060) may not lead the pack, they remain competitive, especially when compared to its predecessors and other models like GPT-4 and LLaMA2. The balanced performance across these metrics highlights GAIA-70b's versatility and its potential to enhance operation monitoring systems by providing insights that are not only safe but also logical, factual, and stable.

As shown in \Cref{table_monitoring} and \Cref{fig_eval_operationmonitoring}, the GAIA-70b model distinguishes itself primarily through its unparalleled commitment to safety, achieving an exceptional score of 9.806 in operation monitoring, the highest among its peers. This preeminence in safety underscores GAIA-70b's capability to deliver reliable outputs while significantly reducing risks, marking it as the optimal choice for applications where safety is paramount.

% ---------

\subsubsection{Evaluation on Power System General Knowledge}

\begin{table}[h!]
\centering
\caption{Evaluation on Power System General Knowledge}
\small
\label{table_general}
\begin{tabular}{lcccc}
\toprule
 & \textbf{Factuality} & \textbf{Logicality} & \textbf{Safety} & \textbf{Stability} \\
\midrule
GPT-4 & \textbf{9.498} & \textbf{9.714} & 9.278 & \textbf{8.650} \\
GPT-3.5 & 8.245 & 8.372 & 5.556 & 8.328 \\
LLaMA2-70b & 7.952 & 7.873 & 9.194 & 8.230 \\
LLaMA2-13b & 8.230 & 7.132 & 8.792 & 6.689 \\
LLaMA2-7b & 6.977 & 6.826 & 9.500 & 6.459 \\
\textbf{GAIA-70b} & 8.257 & 8.150 & 9.694 & 8.230 \\
\textbf{GAIA-13b} & 5.859 & 8.231 & \textbf{9.750} & 6.720 \\
\textbf{GAIA-7b} & 5.859 & 8.231 & \textbf{9.750} & 6.720 \\
\bottomrule
\end{tabular}
\end{table}

\begin{figure}[h!]
\centering
\includegraphics[width=0.48\textwidth,height=0.48\textwidth]{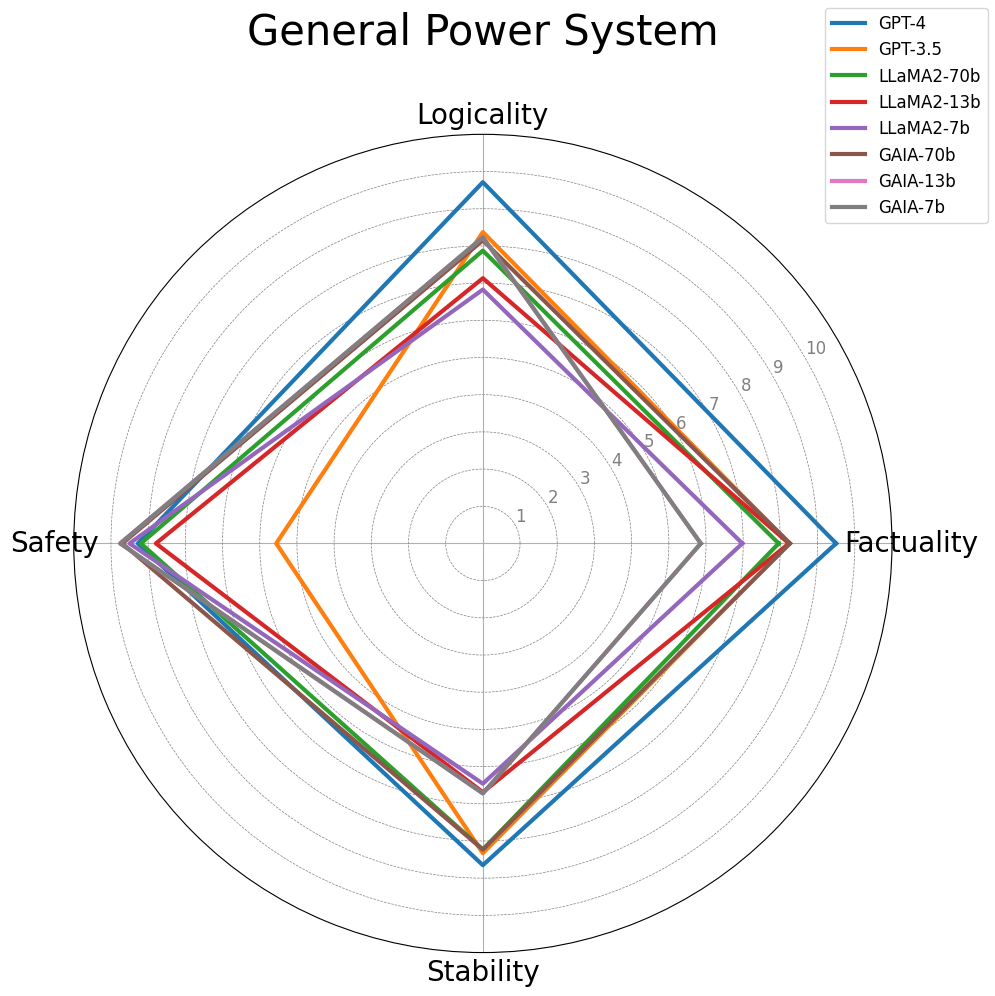}
\caption{Evaluation Result on General Power System} \label{fig_eval_general}
\end{figure}

% In the domain of power system general knowledge, the GAIA-70b model distinguishes itself with robust performance across essential metrics. It achieves a factuality score of 8.257, indicating its ability to provide accurate and reliable information crucial for power system operations. Its logicality score 8.150 ensures that the generated outputs are correct and logically coherent, facilitating practical and actionable insights. Notably, GAIA-70b excels in safety, with a leading score of 9.694, highlighting its exceptional capability to produce safe, guideline-compliant outputs vital for operational safety. Furthermore, its stability score 8.230 demonstrates consistent performance, making GAIA-70b a highly reliable model for applications requiring deep understanding and analysis of power systems.

As demonstrated in \Cref{table_general} and \Cref{fig_eval_general}, the GAIA-70b model establishes a new benchmark in power system general knowledge, achieving an exceptional safety score of 9.694, thereby setting a new standard for operational safety. This achievement highlights GAIA-70b's exceptional ability to generate guideline-compliant outputs and establishes it as the foremost choice for ensuring the highest levels of safety in power system applications.

% ----------

\subsubsection{Evaluation on Power Dispatch}

\begin{table}[h!]
\centering
\caption{Evaluation on Power Dispatch}
\small
\label{table_dispatch}
\begin{tabular}{lcccc}
\toprule
 & \textbf{Factuality} & \textbf{Logicality} & \textbf{Safety} & \textbf{Stability} \\
\midrule
GPT-4 & \textbf{7.419} & \textbf{9.036} & 9.292 & \textbf{8.640} \\
GPT-3.5 & 6.289 & 7.487 & 9.194 & 8.080 \\
LLaMA2-70b & 5.556 & 7.053 & 9.625 & 7.500 \\
LLaMA2-13b & 5.390 & 7.275 & 9.653 & 6.560 \\
LLaMA2-7b & 4.575 & 6.890 & 9.736 & 5.760 \\
\textbf{GAIA-70b} & 5.859 & 8.231 & \textbf{9.750} & 7.900 \\
\textbf{GAIA-13b} & 5.556 & 8.019 & 9.694 & 6.460 \\
\textbf{GAIA-7b} & 4.997 & 7.098 & 9.681 & 5.640 \\
\bottomrule
\end{tabular}
\end{table}

\begin{figure}[h!]
\centering
\includegraphics[width=0.48\textwidth,height=0.48\textwidth]{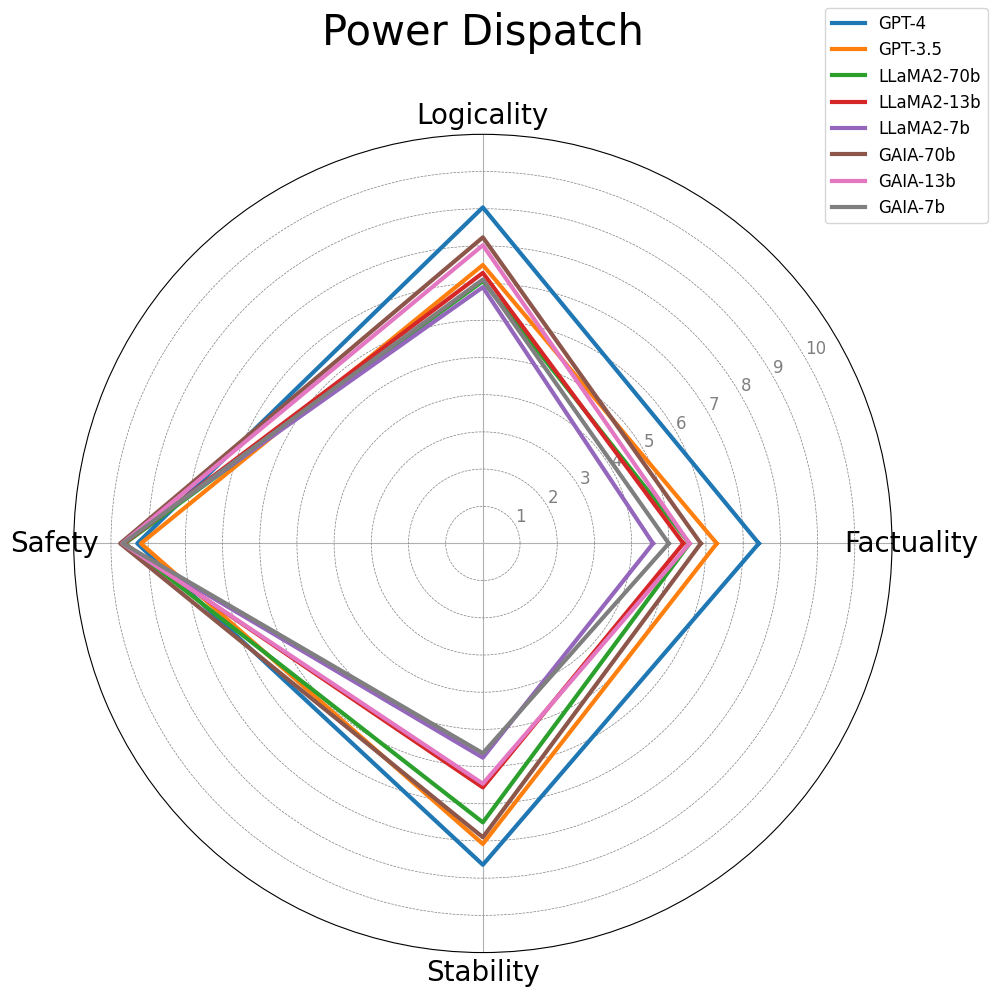}
\caption{Evaluation Result on Power Dispatch} \label{fig_eval_powerdispatch}
\end{figure}

In \Cref{table_dispatch} and \Cref{fig_eval_powerdispatch}, The GAIA-70b model demonstrates a significant advantage in the power dispatch domain, particularly excelling in safety with a score of 9.750, the highest among the models evaluated. This indicates its exceptional ability to generate outputs that adhere to the stringent safety standards required in power system operations. Additionally, its logicality score of 8.231 reflects a solid capability to produce logically coherent and actionable insights, crucial for making informed decisions in power dispatch scenarios. 

% ----------

\subsubsection{Evaluation on Black Start}

\begin{table}[h!]
\centering
\caption{Evaluation on Black Start}
\small
\label{table_black_start}
\begin{tabular}{lcccc}
\toprule
 & \textbf{Factuality} & \textbf{Logicality} & \textbf{Safety} & \textbf{Stability} \\
\midrule
GPT-4 & \textbf{8.394} & \textbf{8.837} & 9.571 & \textbf{8.648} \\
GPT-3.5 & 7.847 & 7.278 & 9.357 & 8.544 \\
LLaMA2-70b & 6.098 & 7.530 & 9.460 & 7.469 \\
LLaMA2-13b & 6.260 & 7.002 & 9.452 & 7.718 \\
LLaMA2-7b & 4.706 & 4.916 & \textbf{9.611} & 6.262 \\
\textbf{GAIA-70b} & 8.313 & 7.662 & 9.508 & 7.673 \\
\textbf{GAIA-13b} & 7.166 & 6.931 & 9.071 & 7.118 \\
\textbf{GAIA-7b} & 7.329 & 5.657 & 9.571 & 7.086 \\
\bottomrule
\end{tabular}
\end{table}

\begin{figure}[h!]
\centering
\includegraphics[width=0.48\textwidth,height=0.48\textwidth]{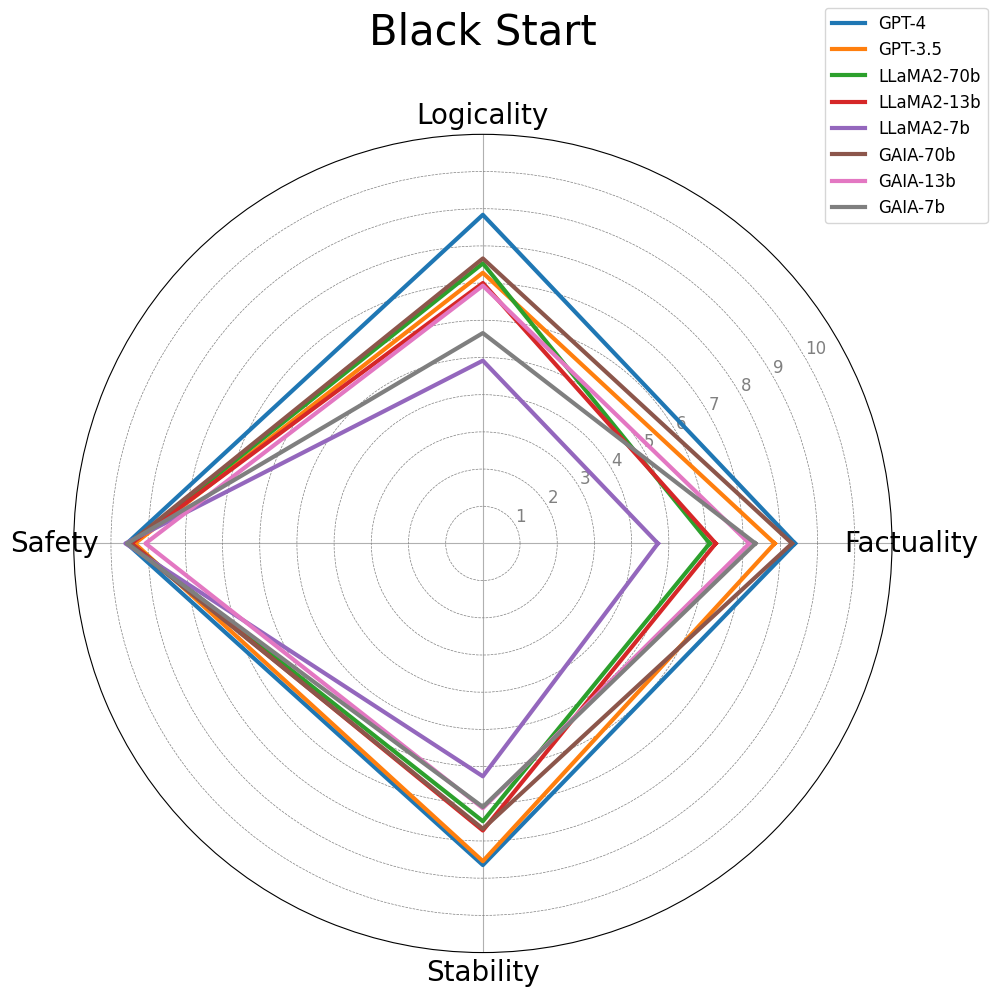}
\caption{Evaluation Result on Black Start} \label{fig_eval_blackstart}
\end{figure}

As shown in \Cref{table_black_start} and \Cref{fig_eval_blackstart}, the GAIA-70b model exhibits commendable performance across various critical metrics in the context of black start procedures, as highlighted in the evaluation. With a factuality score of 8.313, GAIA-70b closely approaches the leading score, showcasing its ability to generate accurate and reliable information essential during the complex process of restoring power after a blackout. Although its logicality score of 7.662 is not the highest, it still reflects a strong capacity for producing coherent and logically structured outputs, a vital attribute for formulating effective black start strategies. The model also achieves a safety score of 9.508, indicating high adherence to safety protocols and standards, which is paramount in the high-stakes environment of black start operations. Furthermore, with a stability score of 7.673, GAIA-70b is a consistent and reliable tool in the dynamic and unpredictable scenarios often encountered during black start procedures. 

%% file: cas-dc-template.bbl
\begin{thebibliography}{50}
\expandafter\ifx\csname natexlab\endcsname\relax\def\natexlab#1{#1}\fi
\providecommand{\url}[1]{\texttt{#1}}
\providecommand{\href}[2]{#2}
\providecommand{\path}[1]{#1}
\providecommand{\DOIprefix}{doi:}
\providecommand{\ArXivprefix}{arXiv:}
\providecommand{\URLprefix}{URL: }
\providecommand{\Pubmedprefix}{pmid:}
\providecommand{\doi}[1]{\href{http://dx.doi.org/#1}{\path{#1}}}
\providecommand{\Pubmed}[1]{\href{pmid:#1}{\path{#1}}}
\providecommand{\bibinfo}[2]{#2}
\ifx\xfnm\relax \def\xfnm[#1]{\unskip,\space#1}\fi
%Type = Article
\bibitem[{Achiam et~al.(2023)Achiam, Adler, Agarwal, Ahmad, Akkaya, Aleman, Almeida, Altenschmidt, Altman, Anadkat et~al.}]{openai2023gpt4}
\bibinfo{author}{Achiam, J.}, \bibinfo{author}{Adler, S.}, \bibinfo{author}{Agarwal, S.}, \bibinfo{author}{Ahmad, L.}, \bibinfo{author}{Akkaya, I.}, \bibinfo{author}{Aleman, F.L.}, \bibinfo{author}{Almeida, D.}, \bibinfo{author}{Altenschmidt, J.}, \bibinfo{author}{Altman, S.}, \bibinfo{author}{Anadkat, S.}, et~al., \bibinfo{year}{2023}.
\newblock \bibinfo{title}{Gpt-4 technical report}.
\newblock \bibinfo{journal}{arXiv preprint arXiv:2303.08774} .
%Type = Misc
\bibitem[{Ainslie et~al.(2023)Ainslie, Lee-Thorp, de~Jong, Zemlyanskiy, Lebrón and Sanghai}]{ainslie2023gqa}
\bibinfo{author}{Ainslie, J.}, \bibinfo{author}{Lee-Thorp, J.}, \bibinfo{author}{de~Jong, M.}, \bibinfo{author}{Zemlyanskiy, Y.}, \bibinfo{author}{Lebrón, F.}, \bibinfo{author}{Sanghai, S.}, \bibinfo{year}{2023}.
\newblock \bibinfo{title}{Gqa: Training generalized multi-query transformer models from multi-head checkpoints}.
\newblock \href{http://arxiv.org/abs/2305.13245}{\tt arXiv:2305.13245}.
%Type = Article
\bibitem[{Almazrouei et~al.(2023)Almazrouei, Alobeidli, Alshamsi, Cappelli, Cojocaru, Debbah, Goffinet, Hesslow, Launay, Malartic et~al.}]{almazrouei2023falcon}
\bibinfo{author}{Almazrouei, E.}, \bibinfo{author}{Alobeidli, H.}, \bibinfo{author}{Alshamsi, A.}, \bibinfo{author}{Cappelli, A.}, \bibinfo{author}{Cojocaru, R.}, \bibinfo{author}{Debbah, M.}, \bibinfo{author}{Goffinet, {\'E}.}, \bibinfo{author}{Hesslow, D.}, \bibinfo{author}{Launay, J.}, \bibinfo{author}{Malartic, Q.}, et~al., \bibinfo{year}{2023}.
\newblock \bibinfo{title}{The falcon series of open language models}.
\newblock \bibinfo{journal}{arXiv preprint arXiv:2311.16867} .
%Type = Book
\bibitem[{Bevrani et~al.(2014)Bevrani, Watanabe and Mitani}]{bevrani2014power}
\bibinfo{author}{Bevrani, H.}, \bibinfo{author}{Watanabe, M.}, \bibinfo{author}{Mitani, Y.}, \bibinfo{year}{2014}.
\newblock \bibinfo{title}{Power system monitoring and control}.
\newblock \bibinfo{publisher}{John Wiley \& Sons}.
%Type = Article
\bibitem[{Blecher et~al.(2023)Blecher, Cucurull, Scialom and Stojnic}]{blecher2023nougat}
\bibinfo{author}{Blecher, L.}, \bibinfo{author}{Cucurull, G.}, \bibinfo{author}{Scialom, T.}, \bibinfo{author}{Stojnic, R.}, \bibinfo{year}{2023}.
\newblock \bibinfo{title}{Nougat: Neural optical understanding for academic documents}.
\newblock \bibinfo{journal}{arXiv preprint arXiv:2308.13418} .
%Type = Article
\bibitem[{Cao et~al.(2024)Cao, Zhao, Cheng, Shu, Liu, Liang, Zhao and Li}]{cao2024survey}
\bibinfo{author}{Cao, Y.}, \bibinfo{author}{Zhao, H.}, \bibinfo{author}{Cheng, Y.}, \bibinfo{author}{Shu, T.}, \bibinfo{author}{Liu, G.}, \bibinfo{author}{Liang, G.}, \bibinfo{author}{Zhao, J.}, \bibinfo{author}{Li, Y.}, \bibinfo{year}{2024}.
\newblock \bibinfo{title}{Survey on large language model-enhanced reinforcement learning: Concept, taxonomy, and methods}.
\newblock \bibinfo{journal}{arXiv preprint arXiv:2404.00282} .
%Type = Article
\bibitem[{Chowdhury and Rahman(1990)}]{chowdhury1990review}
\bibinfo{author}{Chowdhury, B.H.}, \bibinfo{author}{Rahman, S.}, \bibinfo{year}{1990}.
\newblock \bibinfo{title}{A review of recent advances in economic dispatch}.
\newblock \bibinfo{journal}{IEEE transactions on power systems} \bibinfo{volume}{5}, \bibinfo{pages}{1248--1259}.
%Type = Article
\bibitem[{Chung et~al.(2022)Chung, Hou, Longpre, Zoph, Tay, Fedus, Li, Wang, Dehghani, Brahma et~al.}]{chung2022scaling}
\bibinfo{author}{Chung, H.W.}, \bibinfo{author}{Hou, L.}, \bibinfo{author}{Longpre, S.}, \bibinfo{author}{Zoph, B.}, \bibinfo{author}{Tay, Y.}, \bibinfo{author}{Fedus, W.}, \bibinfo{author}{Li, Y.}, \bibinfo{author}{Wang, X.}, \bibinfo{author}{Dehghani, M.}, \bibinfo{author}{Brahma, S.}, et~al., \bibinfo{year}{2022}.
\newblock \bibinfo{title}{Scaling instruction-finetuned language models}.
\newblock \bibinfo{journal}{arXiv preprint arXiv:2210.11416} .
%Type = Article
\bibitem[{Cordonnier et~al.(2020)Cordonnier, Loukas and Jaggi}]{cordonnier2020multi}
\bibinfo{author}{Cordonnier, J.B.}, \bibinfo{author}{Loukas, A.}, \bibinfo{author}{Jaggi, M.}, \bibinfo{year}{2020}.
\newblock \bibinfo{title}{Multi-head attention: Collaborate instead of concatenate}.
\newblock \bibinfo{journal}{arXiv preprint arXiv:2006.16362} .
%Type = Misc
\bibitem[{Dettmers et~al.(2023)Dettmers, Pagnoni, Holtzman and Zettlemoyer}]{dettmers2023qlora}
\bibinfo{author}{Dettmers, T.}, \bibinfo{author}{Pagnoni, A.}, \bibinfo{author}{Holtzman, A.}, \bibinfo{author}{Zettlemoyer, L.}, \bibinfo{year}{2023}.
\newblock \bibinfo{title}{Qlora: Efficient finetuning of quantized llms}.
\newblock \href{http://arxiv.org/abs/2305.14314}{\tt arXiv:2305.14314}.
%Type = Misc
\bibitem[{Fanqi et~al.(2023)Fanqi, Xinting, Tao, Xiaojun, Wei and Shuming}]{wan2023explore}
\bibinfo{author}{Fanqi, W.}, \bibinfo{author}{Xinting, H.}, \bibinfo{author}{Tao, Y.}, \bibinfo{author}{Xiaojun, Q.}, \bibinfo{author}{Wei, B.}, \bibinfo{author}{Shuming, S.}, \bibinfo{year}{2023}.
\newblock \bibinfo{title}{Explore-instruct: Enhancing domain-specific instruction coverage through active exploration}.
\newblock \href{http://arxiv.org/abs/2310.09168}{\tt arXiv:2310.09168}.
%Type = Article
\bibitem[{Gao et~al.(2015)Gao, Cecati and Ding}]{gao2015survey}
\bibinfo{author}{Gao, Z.}, \bibinfo{author}{Cecati, C.}, \bibinfo{author}{Ding, S.X.}, \bibinfo{year}{2015}.
\newblock \bibinfo{title}{A survey of fault diagnosis and fault-tolerant techniques—part i: Fault diagnosis with model-based and signal-based approaches}.
\newblock \bibinfo{journal}{IEEE transactions on industrial electronics} \bibinfo{volume}{62}, \bibinfo{pages}{3757--3767}.
%Type = Article
\bibitem[{Glavic et~al.(2017)Glavic, Fonteneau and Ernst}]{glavic2017reinforcement}
\bibinfo{author}{Glavic, M.}, \bibinfo{author}{Fonteneau, R.}, \bibinfo{author}{Ernst, D.}, \bibinfo{year}{2017}.
\newblock \bibinfo{title}{Reinforcement learning for electric power system decision and control: Past considerations and perspectives}.
\newblock \bibinfo{journal}{IFAC-PapersOnLine} \bibinfo{volume}{50}, \bibinfo{pages}{6918--6927}.
%Type = Article
\bibitem[{Gungor et~al.(2012)Gungor, Sahin, Kocak, Ergut, Buccella, Cecati and Hancke}]{gungor2012survey}
\bibinfo{author}{Gungor, V.C.}, \bibinfo{author}{Sahin, D.}, \bibinfo{author}{Kocak, T.}, \bibinfo{author}{Ergut, S.}, \bibinfo{author}{Buccella, C.}, \bibinfo{author}{Cecati, C.}, \bibinfo{author}{Hancke, G.P.}, \bibinfo{year}{2012}.
\newblock \bibinfo{title}{A survey on smart grid potential applications and communication requirements}.
\newblock \bibinfo{journal}{IEEE Transactions on industrial informatics} \bibinfo{volume}{9}, \bibinfo{pages}{28--42}.
%Type = Article
\bibitem[{Hastings(1970)}]{hastings1970monte}
\bibinfo{author}{Hastings, W.K.}, \bibinfo{year}{1970}.
\newblock \bibinfo{title}{Monte carlo sampling methods using markov chains and their applications} .
%Type = Misc
\bibitem[{Hendrycks et~al.(2021a)Hendrycks, Burns, Basart, Zou, Mazeika, Song and Steinhardt}]{hendrycks2021measuring}
\bibinfo{author}{Hendrycks, D.}, \bibinfo{author}{Burns, C.}, \bibinfo{author}{Basart, S.}, \bibinfo{author}{Zou, A.}, \bibinfo{author}{Mazeika, M.}, \bibinfo{author}{Song, D.}, \bibinfo{author}{Steinhardt, J.}, \bibinfo{year}{2021}a.
\newblock \bibinfo{title}{Measuring massive multitask language understanding}.
\newblock \href{http://arxiv.org/abs/2009.03300}{\tt arXiv:2009.03300}.
%Type = Misc
\bibitem[{Hendrycks et~al.(2021b)Hendrycks, Burns, Kadavath, Arora, Basart, Tang, Song and Steinhardt}]{hendrycks2021measuringmath}
\bibinfo{author}{Hendrycks, D.}, \bibinfo{author}{Burns, C.}, \bibinfo{author}{Kadavath, S.}, \bibinfo{author}{Arora, A.}, \bibinfo{author}{Basart, S.}, \bibinfo{author}{Tang, E.}, \bibinfo{author}{Song, D.}, \bibinfo{author}{Steinhardt, J.}, \bibinfo{year}{2021}b.
\newblock \bibinfo{title}{Measuring mathematical problem solving with the math dataset}.
\newblock \href{http://arxiv.org/abs/2103.03874}{\tt arXiv:2103.03874}.
%Type = Misc
\bibitem[{Houlsby et~al.(2019)Houlsby, Giurgiu, Jastrzebski, Morrone, de~Laroussilhe, Gesmundo, Attariyan and Gelly}]{houlsby2019parameterefficient}
\bibinfo{author}{Houlsby, N.}, \bibinfo{author}{Giurgiu, A.}, \bibinfo{author}{Jastrzebski, S.}, \bibinfo{author}{Morrone, B.}, \bibinfo{author}{de~Laroussilhe, Q.}, \bibinfo{author}{Gesmundo, A.}, \bibinfo{author}{Attariyan, M.}, \bibinfo{author}{Gelly, S.}, \bibinfo{year}{2019}.
\newblock \bibinfo{title}{Parameter-efficient transfer learning for nlp}.
\newblock \href{http://arxiv.org/abs/1902.00751}{\tt arXiv:1902.00751}.
%Type = Misc
\bibitem[{Hu et~al.(2021)Hu, Shen, Wallis, Allen-Zhu, Li, Wang, Wang and Chen}]{hu2021lora}
\bibinfo{author}{Hu, E.J.}, \bibinfo{author}{Shen, Y.}, \bibinfo{author}{Wallis, P.}, \bibinfo{author}{Allen-Zhu, Z.}, \bibinfo{author}{Li, Y.}, \bibinfo{author}{Wang, S.}, \bibinfo{author}{Wang, L.}, \bibinfo{author}{Chen, W.}, \bibinfo{year}{2021}.
\newblock \bibinfo{title}{Lora: Low-rank adaptation of large language models}.
\newblock \href{http://arxiv.org/abs/2106.09685}{\tt arXiv:2106.09685}.
%Type = Article
\bibitem[{Lee et~al.(1985)Lee, Park and Ortiz}]{lee1985united}
\bibinfo{author}{Lee, K.}, \bibinfo{author}{Park, Y.}, \bibinfo{author}{Ortiz, J.}, \bibinfo{year}{1985}.
\newblock \bibinfo{title}{A united approach to optimal real and reactive power dispatch}.
\newblock \bibinfo{journal}{IEEE Transactions on power Apparatus and systems} , \bibinfo{pages}{1147--1153}.
%Type = Article
\bibitem[{Liu et~al.(2024)Liu, Bai, Wen, Wang, Liu, Liang, Zhao and Dong}]{liulfllm}
\bibinfo{author}{Liu, G.}, \bibinfo{author}{Bai, Y.}, \bibinfo{author}{Wen, K.}, \bibinfo{author}{Wang, X.}, \bibinfo{author}{Liu, Y.}, \bibinfo{author}{Liang, G.}, \bibinfo{author}{Zhao, J.}, \bibinfo{author}{Dong, Z.Y.}, \bibinfo{year}{2024}.
\newblock \bibinfo{title}{Lfllm: A large language model for load forecasting} .
%Type = Article
\bibitem[{Liu et~al.(2023a)Liu, Yuan, Fu, Jiang, Hayashi and Neubig}]{liu2023pre}
\bibinfo{author}{Liu, P.}, \bibinfo{author}{Yuan, W.}, \bibinfo{author}{Fu, J.}, \bibinfo{author}{Jiang, Z.}, \bibinfo{author}{Hayashi, H.}, \bibinfo{author}{Neubig, G.}, \bibinfo{year}{2023}a.
\newblock \bibinfo{title}{Pre-train, prompt, and predict: A systematic survey of prompting methods in natural language processing}.
\newblock \bibinfo{journal}{ACM Computing Surveys} \bibinfo{volume}{55}, \bibinfo{pages}{1--35}.
%Type = Misc
\bibitem[{Liu et~al.(2023b)Liu, Zheng, Du, Ding, Qian, Yang and Tang}]{liu2023gpt}
\bibinfo{author}{Liu, X.}, \bibinfo{author}{Zheng, Y.}, \bibinfo{author}{Du, Z.}, \bibinfo{author}{Ding, M.}, \bibinfo{author}{Qian, Y.}, \bibinfo{author}{Yang, Z.}, \bibinfo{author}{Tang, J.}, \bibinfo{year}{2023}b.
\newblock \bibinfo{title}{Gpt understands, too}.
\newblock \href{http://arxiv.org/abs/2103.10385}{\tt arXiv:2103.10385}.
%Type = Misc
\bibitem[{Luo et~al.(2023)Luo, Sun, Xu, Zhao, Lou, Tao, Geng, Lin, Chen and Zhang}]{luo2023wizardmath}
\bibinfo{author}{Luo, H.}, \bibinfo{author}{Sun, Q.}, \bibinfo{author}{Xu, C.}, \bibinfo{author}{Zhao, P.}, \bibinfo{author}{Lou, J.}, \bibinfo{author}{Tao, C.}, \bibinfo{author}{Geng, X.}, \bibinfo{author}{Lin, Q.}, \bibinfo{author}{Chen, S.}, \bibinfo{author}{Zhang, D.}, \bibinfo{year}{2023}.
\newblock \bibinfo{title}{Wizardmath: Empowering mathematical reasoning for large language models via reinforced evol-instruct}.
\newblock \href{http://arxiv.org/abs/2308.09583}{\tt arXiv:2308.09583}.
%Type = Inproceedings
\bibitem[{Ma et~al.(2014)Ma, Guo, Su, Chen, Du, Yang, Li, Lin and Geng}]{ma2014deep}
\bibinfo{author}{Ma, Y.}, \bibinfo{author}{Guo, Z.}, \bibinfo{author}{Su, J.}, \bibinfo{author}{Chen, Y.}, \bibinfo{author}{Du, X.}, \bibinfo{author}{Yang, Y.}, \bibinfo{author}{Li, C.}, \bibinfo{author}{Lin, Y.}, \bibinfo{author}{Geng, Y.}, \bibinfo{year}{2014}.
\newblock \bibinfo{title}{Deep learning for fault diagnosis based on multi-sourced heterogeneous data}, in: \bibinfo{booktitle}{2014 International Conference on Power System Technology}, \bibinfo{organization}{IEEE}. pp. \bibinfo{pages}{740--745}.
%Type = Misc
\bibitem[{OpenAI(2022)}]{chatgpt}
\bibinfo{author}{OpenAI}, \bibinfo{year}{2022}.
\newblock \bibinfo{title}{{C}hat{G}{P}{T}}.
\newblock \bibinfo{howpublished}{\url{http://chat.openai.com/}}.
%Type = Article
\bibitem[{Ouyang et~al.(2022)Ouyang, Wu, Jiang, Almeida, Wainwright, Mishkin, Zhang, Agarwal, Slama, Ray et~al.}]{ouyang2022training}
\bibinfo{author}{Ouyang, L.}, \bibinfo{author}{Wu, J.}, \bibinfo{author}{Jiang, X.}, \bibinfo{author}{Almeida, D.}, \bibinfo{author}{Wainwright, C.}, \bibinfo{author}{Mishkin, P.}, \bibinfo{author}{Zhang, C.}, \bibinfo{author}{Agarwal, S.}, \bibinfo{author}{Slama, K.}, \bibinfo{author}{Ray, A.}, et~al., \bibinfo{year}{2022}.
\newblock \bibinfo{title}{Training language models to follow instructions with human feedback}.
\newblock \bibinfo{journal}{Advances in Neural Information Processing Systems} \bibinfo{volume}{35}, \bibinfo{pages}{27730--27744}.
%Type = Article
\bibitem[{Patsakis et~al.(2018)Patsakis, Rajan, Aravena, Rios and Oren}]{patsakis2018optimal}
\bibinfo{author}{Patsakis, G.}, \bibinfo{author}{Rajan, D.}, \bibinfo{author}{Aravena, I.}, \bibinfo{author}{Rios, J.}, \bibinfo{author}{Oren, S.}, \bibinfo{year}{2018}.
\newblock \bibinfo{title}{Optimal black start allocation for power system restoration}.
\newblock \bibinfo{journal}{IEEE Transactions on Power Systems} \bibinfo{volume}{33}, \bibinfo{pages}{6766--6776}.
%Type = Article
\bibitem[{Roziere et~al.(2023)Roziere, Gehring, Gloeckle, Sootla, Gat, Tan, Adi, Liu, Remez, Rapin et~al.}]{roziere2023code}
\bibinfo{author}{Roziere, B.}, \bibinfo{author}{Gehring, J.}, \bibinfo{author}{Gloeckle, F.}, \bibinfo{author}{Sootla, S.}, \bibinfo{author}{Gat, I.}, \bibinfo{author}{Tan, X.E.}, \bibinfo{author}{Adi, Y.}, \bibinfo{author}{Liu, J.}, \bibinfo{author}{Remez, T.}, \bibinfo{author}{Rapin, J.}, et~al., \bibinfo{year}{2023}.
\newblock \bibinfo{title}{Code llama: Open foundation models for code}.
\newblock \bibinfo{journal}{arXiv preprint arXiv:2308.12950} .
%Type = Misc
\bibitem[{Schulman et~al.(2017)Schulman, Wolski, Dhariwal, Radford and Klimov}]{schulman2017proximal}
\bibinfo{author}{Schulman, J.}, \bibinfo{author}{Wolski, F.}, \bibinfo{author}{Dhariwal, P.}, \bibinfo{author}{Radford, A.}, \bibinfo{author}{Klimov, O.}, \bibinfo{year}{2017}.
\newblock \bibinfo{title}{Proximal policy optimization algorithms}.
\newblock \href{http://arxiv.org/abs/1707.06347}{\tt arXiv:1707.06347}.
%Type = Article
\bibitem[{Sener and Savarese(2017)}]{sener2017active}
\bibinfo{author}{Sener, O.}, \bibinfo{author}{Savarese, S.}, \bibinfo{year}{2017}.
\newblock \bibinfo{title}{Active learning for convolutional neural networks: A core-set approach}.
\newblock \bibinfo{journal}{arXiv preprint arXiv:1708.00489} .
%Type = Misc
\bibitem[{Shazeer(2019)}]{shazeer2019fast}
\bibinfo{author}{Shazeer, N.}, \bibinfo{year}{2019}.
\newblock \bibinfo{title}{Fast transformer decoding: One write-head is all you need}.
\newblock \href{http://arxiv.org/abs/1911.02150}{\tt arXiv:1911.02150}.
%Type = Article
\bibitem[{Shazeer(2020)}]{shazeer2020glu}
\bibinfo{author}{Shazeer, N.}, \bibinfo{year}{2020}.
\newblock \bibinfo{title}{Glu variants improve transformer}.
\newblock \bibinfo{journal}{arXiv preprint arXiv:2002.05202} .
%Type = Article
\bibitem[{Singhal et~al.(2023)Singhal, Azizi, Tu, Mahdavi, Wei, Chung, Scales, Tanwani, Cole-Lewis, Pfohl et~al.}]{singhal2023large}
\bibinfo{author}{Singhal, K.}, \bibinfo{author}{Azizi, S.}, \bibinfo{author}{Tu, T.}, \bibinfo{author}{Mahdavi, S.S.}, \bibinfo{author}{Wei, J.}, \bibinfo{author}{Chung, H.W.}, \bibinfo{author}{Scales, N.}, \bibinfo{author}{Tanwani, A.}, \bibinfo{author}{Cole-Lewis, H.}, \bibinfo{author}{Pfohl, S.}, et~al., \bibinfo{year}{2023}.
\newblock \bibinfo{title}{Large language models encode clinical knowledge}.
\newblock \bibinfo{journal}{Nature} \bibinfo{volume}{620}, \bibinfo{pages}{172--180}.
%Type = Misc
\bibitem[{Su et~al.(2023)Su, Lu, Pan, Murtadha, Wen and Liu}]{su2023roformer}
\bibinfo{author}{Su, J.}, \bibinfo{author}{Lu, Y.}, \bibinfo{author}{Pan, S.}, \bibinfo{author}{Murtadha, A.}, \bibinfo{author}{Wen, B.}, \bibinfo{author}{Liu, Y.}, \bibinfo{year}{2023}.
\newblock \bibinfo{title}{Roformer: Enhanced transformer with rotary position embedding}.
\newblock \href{http://arxiv.org/abs/2104.09864}{\tt arXiv:2104.09864}.
%Type = Misc
\bibitem[{Team(2023)}]{MosaicML2023Introducing}
\bibinfo{author}{Team, M.N.}, \bibinfo{year}{2023}.
\newblock \bibinfo{title}{Introducing mpt-7b: A new standard for open-source, commercially usable llms}.
\newblock \URLprefix \url{www.mosaicml.com/blog/mpt-7b}. \bibinfo{note}{accessed: 2023-05-05}.
%Type = Misc
\bibitem[{Touvron et~al.(2023a)Touvron, Lavril, Izacard, Martinet, Lachaux, Lacroix, Rozière, Goyal, Hambro, Azhar, Rodriguez, Joulin, Grave and Lample}]{touvron2023llama}
\bibinfo{author}{Touvron, H.}, \bibinfo{author}{Lavril, T.}, \bibinfo{author}{Izacard, G.}, \bibinfo{author}{Martinet, X.}, \bibinfo{author}{Lachaux, M.A.}, \bibinfo{author}{Lacroix, T.}, \bibinfo{author}{Rozière, B.}, \bibinfo{author}{Goyal, N.}, \bibinfo{author}{Hambro, E.}, \bibinfo{author}{Azhar, F.}, \bibinfo{author}{Rodriguez, A.}, \bibinfo{author}{Joulin, A.}, \bibinfo{author}{Grave, E.}, \bibinfo{author}{Lample, G.}, \bibinfo{year}{2023}a.
\newblock \bibinfo{title}{Llama: Open and efficient foundation language models}.
\newblock \href{http://arxiv.org/abs/2302.13971}{\tt arXiv:2302.13971}.
%Type = Article
\bibitem[{Touvron et~al.(2023b)Touvron, Martin, Stone, Albert, Almahairi, Babaei, Bashlykov, Batra, Bhargava, Bhosale et~al.}]{touvron2023llama2}
\bibinfo{author}{Touvron, H.}, \bibinfo{author}{Martin, L.}, \bibinfo{author}{Stone, K.}, \bibinfo{author}{Albert, P.}, \bibinfo{author}{Almahairi, A.}, \bibinfo{author}{Babaei, Y.}, \bibinfo{author}{Bashlykov, N.}, \bibinfo{author}{Batra, S.}, \bibinfo{author}{Bhargava, P.}, \bibinfo{author}{Bhosale, S.}, et~al., \bibinfo{year}{2023}b.
\newblock \bibinfo{title}{Llama 2: Open foundation and fine-tuned chat models}.
\newblock \bibinfo{journal}{arXiv preprint arXiv:2307.09288} .
%Type = Article
\bibitem[{Valinejad et~al.(2023)Valinejad, Mili, Yu, Van Der~Wal and Xu}]{valinejad2023computational}
\bibinfo{author}{Valinejad, J.}, \bibinfo{author}{Mili, L.}, \bibinfo{author}{Yu, X.}, \bibinfo{author}{Van Der~Wal, C.N.}, \bibinfo{author}{Xu, Y.}, \bibinfo{year}{2023}.
\newblock \bibinfo{title}{Computational social science in smart power systems: Reliability, resilience, and restoration}.
\newblock \bibinfo{journal}{Energy Conversion and Economics} \bibinfo{volume}{4}, \bibinfo{pages}{159--170}.
%Type = Misc
\bibitem[{Wang et~al.(2023)Wang, Kordi, Mishra, Liu, Smith, Khashabi and Hajishirzi}]{wang2023selfinstruct}
\bibinfo{author}{Wang, Y.}, \bibinfo{author}{Kordi, Y.}, \bibinfo{author}{Mishra, S.}, \bibinfo{author}{Liu, A.}, \bibinfo{author}{Smith, N.A.}, \bibinfo{author}{Khashabi, D.}, \bibinfo{author}{Hajishirzi, H.}, \bibinfo{year}{2023}.
\newblock \bibinfo{title}{Self-instruct: Aligning language models with self-generated instructions}.
\newblock \href{http://arxiv.org/abs/2212.10560}{\tt arXiv:2212.10560}.
%Type = Misc
\bibitem[{Wei et~al.(2023)Wei, Wang, Schuurmans, Bosma, Ichter, Xia, Chi, Le and Zhou}]{wei2023chainofthought}
\bibinfo{author}{Wei, J.}, \bibinfo{author}{Wang, X.}, \bibinfo{author}{Schuurmans, D.}, \bibinfo{author}{Bosma, M.}, \bibinfo{author}{Ichter, B.}, \bibinfo{author}{Xia, F.}, \bibinfo{author}{Chi, E.}, \bibinfo{author}{Le, Q.}, \bibinfo{author}{Zhou, D.}, \bibinfo{year}{2023}.
\newblock \bibinfo{title}{Chain-of-thought prompting elicits reasoning in large language models}.
\newblock \href{http://arxiv.org/abs/2201.11903}{\tt arXiv:2201.11903}.
%Type = Article
\bibitem[{Wen et~al.(2019)Wen, Zhou, Yang and Lu}]{wen2019optimal}
\bibinfo{author}{Wen, L.}, \bibinfo{author}{Zhou, K.}, \bibinfo{author}{Yang, S.}, \bibinfo{author}{Lu, X.}, \bibinfo{year}{2019}.
\newblock \bibinfo{title}{Optimal load dispatch of community microgrid with deep learning based solar power and load forecasting}.
\newblock \bibinfo{journal}{Energy} \bibinfo{volume}{171}, \bibinfo{pages}{1053--1065}.
%Type = Book
\bibitem[{Wood et~al.(2013)Wood, Wollenberg and Shebl{\'e}}]{wood2013power}
\bibinfo{author}{Wood, A.J.}, \bibinfo{author}{Wollenberg, B.F.}, \bibinfo{author}{Shebl{\'e}, G.B.}, \bibinfo{year}{2013}.
\newblock \bibinfo{title}{Power generation, operation, and control}.
\newblock \bibinfo{publisher}{John Wiley \& Sons}.
%Type = Article
\bibitem[{Xie et~al.(2023)Xie, Wan, Huang, Yin, Liu, Wang, Linghu, Kit, Grazian, Zhang et~al.}]{xie2023darwin}
\bibinfo{author}{Xie, T.}, \bibinfo{author}{Wan, Y.}, \bibinfo{author}{Huang, W.}, \bibinfo{author}{Yin, Z.}, \bibinfo{author}{Liu, Y.}, \bibinfo{author}{Wang, S.}, \bibinfo{author}{Linghu, Q.}, \bibinfo{author}{Kit, C.}, \bibinfo{author}{Grazian, C.}, \bibinfo{author}{Zhang, W.}, et~al., \bibinfo{year}{2023}.
\newblock \bibinfo{title}{Darwin series: Domain specific large language models for natural science}.
\newblock \bibinfo{journal}{arXiv preprint arXiv:2308.13565} .
%Type = Article
\bibitem[{Xu et~al.(2023a)Xu, Sun, Zheng, Geng, Zhao, Feng, Tao and Jiang}]{xu2023wizardlm}
\bibinfo{author}{Xu, C.}, \bibinfo{author}{Sun, Q.}, \bibinfo{author}{Zheng, K.}, \bibinfo{author}{Geng, X.}, \bibinfo{author}{Zhao, P.}, \bibinfo{author}{Feng, J.}, \bibinfo{author}{Tao, C.}, \bibinfo{author}{Jiang, D.}, \bibinfo{year}{2023}a.
\newblock \bibinfo{title}{Wizardlm: Empowering large language models to follow complex instructions}.
\newblock \bibinfo{journal}{arXiv preprint arXiv:2304.12244} .
%Type = Misc
\bibitem[{Xu et~al.(2023b)Xu, Xie, Qin, Tao and Wang}]{xu2023parameterefficient}
\bibinfo{author}{Xu, L.}, \bibinfo{author}{Xie, H.}, \bibinfo{author}{Qin, S.Z.J.}, \bibinfo{author}{Tao, X.}, \bibinfo{author}{Wang, F.L.}, \bibinfo{year}{2023}b.
\newblock \bibinfo{title}{Parameter-efficient fine-tuning methods for pretrained language models: A critical review and assessment}.
\newblock \href{http://arxiv.org/abs/2312.12148}{\tt arXiv:2312.12148}.
%Type = Article
\bibitem[{Zhang and Sennrich(2019)}]{zhang2019root}
\bibinfo{author}{Zhang, B.}, \bibinfo{author}{Sennrich, R.}, \bibinfo{year}{2019}.
\newblock \bibinfo{title}{Root mean square layer normalization}.
\newblock \bibinfo{journal}{Advances in Neural Information Processing Systems} \bibinfo{volume}{32}.
%Type = Article
\bibitem[{Zhang et~al.(2023)Zhang, Dong, Li, Zhang, Sun, Wang, Li, Hu, Zhang, Wu et~al.}]{zhang2023instruction}
\bibinfo{author}{Zhang, S.}, \bibinfo{author}{Dong, L.}, \bibinfo{author}{Li, X.}, \bibinfo{author}{Zhang, S.}, \bibinfo{author}{Sun, X.}, \bibinfo{author}{Wang, S.}, \bibinfo{author}{Li, J.}, \bibinfo{author}{Hu, R.}, \bibinfo{author}{Zhang, T.}, \bibinfo{author}{Wu, F.}, et~al., \bibinfo{year}{2023}.
\newblock \bibinfo{title}{Instruction tuning for large language models: A survey}.
\newblock \bibinfo{journal}{arXiv preprint arXiv:2308.10792} .
%Type = Article
\bibitem[{Zhao et~al.(2023)Zhao, Zhou, Li, Tang, Wang, Hou, Min, Zhang, Zhang, Dong et~al.}]{zhao2023survey}
\bibinfo{author}{Zhao, W.X.}, \bibinfo{author}{Zhou, K.}, \bibinfo{author}{Li, J.}, \bibinfo{author}{Tang, T.}, \bibinfo{author}{Wang, X.}, \bibinfo{author}{Hou, Y.}, \bibinfo{author}{Min, Y.}, \bibinfo{author}{Zhang, B.}, \bibinfo{author}{Zhang, J.}, \bibinfo{author}{Dong, Z.}, et~al., \bibinfo{year}{2023}.
\newblock \bibinfo{title}{A survey of large language models}.
\newblock \bibinfo{journal}{arXiv preprint arXiv:2303.18223} .
%Type = Article
\bibitem[{Zhou et~al.(2024)Zhou, Zhao, Cheng, Cao, Liang, Liu and Zhao}]{zhouelecbench}
\bibinfo{author}{Zhou, X.}, \bibinfo{author}{Zhao, H.}, \bibinfo{author}{Cheng, Y.}, \bibinfo{author}{Cao, Y.}, \bibinfo{author}{Liang, G.}, \bibinfo{author}{Liu, G.}, \bibinfo{author}{Zhao, J.}, \bibinfo{year}{2024}.
\newblock \bibinfo{title}{Elecbench: A power dispatch evaluation benchmark for large language models}.
\newblock \bibinfo{journal}{arXiv preprint arXiv:2407.05365} .

\end{thebibliography}
